\documentclass[12pt]{revtex4}
\usepackage{color}
\usepackage{epsf}
\usepackage{amssymb}
\usepackage[english]{babel}
\newcommand{\df}{\,\mathrm{d}}
\begin{document}
\sloppy
\begin{center}
{\Large \bf
On the ambiguity of the interfering resonances parameters determination.
}\\[8mm]
A.D.Bukin\\[8mm]
Budker Institute of Nuclear Physics, \\[3mm]
630090, Novosibirsk, Russia\\[8mm]
{\bf Abstract}
\end{center}

In the paper the interfering resonances parameters determination ambiguity is considered.
It is shown that there are two solutions for two fixed width resonances. Analytical relation between different solutions is derived. Numeric experiments for fixed width three and four resonances, and for model energy-dependant width two resonances confirm ambiguity of the resonances parameters determination.

\section{Introduction}
One of the typical tasks during experimental data processing is
determination of the parameters of several resonances
from the experimental cross-section measurements
taking into account their interference with
arbitrary phases.
Often it occurs that for the resonances with arbitrary phases
several almost equally good solutions can be found.
In the paper an attempt to analyze this problem is performed using simple examples.

The preliminary version of this  paper has been published as a preprint of Budker Institute
of Nuclear Physics~\cite{BINP_prep}.

\section{Two resonances interference with simple non-relativistic
Breit-Wigner amplitude
}

Let us consider a model cross section
\begin{equation}
\sigma(E)=\left|\frac{A}{E-m_1+i\frac{\Gamma_1}{2}}+
\frac{B}{E-m_2+i\frac{\Gamma_2}{2}}\right|^2,
\end{equation}
where $A,B$ are some complex numbers (coupling constants), $m_1,m_2,\Gamma_1,\Gamma_2$ are
real numbers (masses and widths of resonances).

First let us derive the conditions which lead to the
identical cross section as a function of energy
with different set of parameters.

Two identical continuous functions should have identical
Fourrier images.

For the function of interest the Fourrier image~\cite{KORN}
is easily calculated:
\begin{equation}
\begin{array}{l}
\phi(t)=\int\limits_{-\infty}^{+\infty}\sigma(E)e^{itE}\df E=
\\[4mm]=
2\pi\left\{
\begin{array}{l}
\left[\frac{A^*A}{\Gamma_1}+
\frac{iA^*B}{m_1-m_2+\frac{i}{2}\left(\Gamma_1+\Gamma_2\right)}\right]
e^{it\left(m_1+\frac{i\Gamma_1}{2}\right)}+
\\ \rule{15mm}{0mm}+
\left[\frac{iB^*A}{m_2-m_1+\frac{i}{2}\left(\Gamma_1+\Gamma_2\right)}+
\frac{B^*B}{\Gamma_2}
\right]e^{it\left(m_2+\frac{i\Gamma_2}{2}\right)},\; t>0,\\[5mm]
\left[\frac{A^*A}{\Gamma_1}-
\frac{iAB^*}{m_1-m_2-\frac{i}{2}\left(\Gamma_1+\Gamma_2\right)}\right]
e^{it\left(m_1-\frac{i\Gamma_1}{2}\right)}+
\\ \rule{15mm}{0mm}+
\left[-\frac{iBA^*}{m_2-m_1-\frac{i}{2}\left(\Gamma_1+\Gamma_2\right)}+
\frac{B^*B}{\Gamma_2}
\right]e^{it\left(m_2-\frac{i\Gamma_2}{2}\right)},\; t<0
\end{array}
\right.
\end{array}
\end{equation}
In order that function
\begin{equation}
\sigma_x(E)=\left|\frac{A_x}{E-m_{1x}+i\frac{\Gamma_{1x}}{2}}+
\frac{B_x}{E-m_{2x}+i\frac{\Gamma_{2x}}{2}}\right|^2
\end{equation}
be equal to
 $\sigma(E)$ at every point  $E$,
 evidently the following equalities should be valid
\begin{equation}
\begin{array}{l}
m_{1x}=m_1, \;\;\;\; \Gamma_{1x}=\Gamma_1, \\[4mm]
\frac{A_x^*A_x}{\Gamma_{1x}}+
\frac{iA_x^*B_x}{m_{1x}-m_{2x}+\frac{i}{2}\left(\Gamma_{1x}+\Gamma_{2x}\right)}=
\frac{A^*A}{\Gamma_1}+
\frac{iA^*B}{m_1-m_2+\frac{i}{2}\left(\Gamma_1+\Gamma_2\right)},\\[4mm]
m_{2x}=m_2, \;\;\; \Gamma_{2x}=\Gamma_2, \\[4mm]
\frac{iB_x^*A_x}{m_{2x}-m_{1x}+\frac{i}{2}\left(\Gamma_{1x}+\Gamma_{2x}\right)}+
\frac{B_x^*B_x}{\Gamma_{2x}}
=
\frac{iB^*A}{m_2-m_1+\frac{i}{2}\left(\Gamma_1+\Gamma_2\right)}+
\frac{B^*B}{\Gamma_2}.
\end{array}
\end{equation}
Apparently the resonance masses should be ordered here otherwise
additional trivial solutions would appear due to parameter sets exchange.

For the amplitudes
 $A_x,B_x$ we have four equations with four variables
 (separate equations for the real and imaginary parts).
Because the equations are non-linear there could be more than one solution.

Since only amplitude absolute value squared has a physical sence
there is a freedom in absolute phases with definite relative phase value.
So we can take, for example, that  $A_x$
is a real number and $B_x$ defines their relative
phase,
or equivalent:
$$
A_x=\left|A_x\right|e^{i\psi},\;B_x=\left|B_x\right|e^{-i\psi},
$$
If the latter definition is admitted then
\begin{equation}
A_x=a_xe^{i\psi_x}, \; B_x=b_xe^{-i\psi_x}, \;
A=ae^{i\psi}, \; B=be^{-i\psi}, \;
\end{equation}
where $a,b,\psi,a_x,b_x,\psi_x$ are real numbers.

Now one can write down system of equations:
\begin{equation}\label{Basic_system}
\left\{
\begin{array}{l}
\frac{a_x^2}{\Gamma_{1}}+\frac{a_xb_x\cdot\left[\frac{\cos(2\psi_x)}{2}
\left(\Gamma_{1}+\Gamma_{2}\right)+\left(m_{1}-m_{2}\right)\sin(2\psi_x)\right]}
{\left(m_{1}-m_{2}\right)^2+\left(\frac{\Gamma_{1}+\Gamma_{2}}{2}\right)^2}=
\\ \rule{20mm}{0mm}=
\frac{a^2}{\Gamma_{1}}+\frac{ab\cdot\left[\frac{\cos(2\psi)}{2}
\left(\Gamma_{1}+\Gamma_{2}\right)+\left(m_{1}-m_{2}\right)\sin(2\psi)\right]}
{\left(m_{1}-m_{2}\right)^2+\left(\frac{\Gamma_{1}+\Gamma_{2}}{2}\right)^2}\\[3mm]
\frac{b_x^2}{\Gamma_{2}}+\frac{a_xb_x\cdot\left[\frac{\cos(2\psi_x)}{2}
\left(\Gamma_{1}+\Gamma_{2}\right)-\left(m_{2}-m_{1}\right)\sin(2\psi_x)\right]}
{\left(m_{1}-m_{2}\right)^2+\left(\frac{\Gamma_{1}+\Gamma_{2}}{2}\right)^2}=
\\ \rule{20mm}{0mm}=
\frac{b^2}{\Gamma_{2}}+\frac{ab\cdot\left[\frac{\cos(2\psi)}{2}
\left(\Gamma_{1}+\Gamma_{2}\right)-\left(m_{2}-m_{1}\right)\sin(2\psi)\right]}
{\left(m_{1}-m_{2}\right)^2+\left(\frac{\Gamma_{1}+\Gamma_{2}}{2}\right)^2}
\\[3mm]
\frac{a_xb_x\cdot\left[\left(m_{1}-m_{2}\right)\cos(2\psi_x)-
\frac{\sin(2\psi_x)}{2}
\left(\Gamma_{1}+\Gamma_{2}\right)\right]}
{\left(m_{1}-m_{2}\right)^2+\left(\frac{\Gamma_{1}+\Gamma_{2}}{2}\right)^2}=
\\ \rule{20mm}{0mm}=
\frac{ab\cdot\left[\left(m_{1}-m_{2}\right)\cos(2\psi)-
\frac{\sin(2\psi)}{2}
\left(\Gamma_{1}+\Gamma_{2}\right)\right]}
{\left(m_{1}-m_{2}\right)^2+\left(\frac{\Gamma_{1}+\Gamma_{2}}{2}\right)^2}\\[3mm]
\frac{a_xb_x\cdot\left[\left(m_{2}-m_{1}\right)\cos(2\psi_x)+
\frac{\sin(2\psi_x)}{2}
\left(\Gamma_{1}+\Gamma_{2}\right)\right]}
{\left(m_{1}-m_{2}\right)^2+\left(\frac{\Gamma_{1}+\Gamma_{2}}{2}\right)^2}=
\\ \rule{20mm}{0mm}=
\frac{ab\cdot\left[\left(m_{2}-m_{1}\right)\cos(2\psi)+
\frac{\sin(2\psi)}{2}
\left(\Gamma_{1}+\Gamma_{2}\right)\right]}
{\left(m_{1}-m_{2}\right)^2+\left(\frac{\Gamma_{1}+\Gamma_{2}}{2}\right)^2}
\end{array}
\right.
\end{equation}
Evidently the last two equations are identical, so
for three unknown variables
 $a_x,b_x,\psi_x$ we have three independent equations.

Trivial solution: $a_x=a,\; b_x=b,\; \psi_x=\psi$.
Let us check whether  there are some other solutions.
First exclude  $\psi_x$.
One equation without $\psi_x$ can be derived by subtracting
the second equation in (\ref{Basic_system}) from the first one:
\begin{equation}
\frac{a_x^2}{\Gamma_1}-\frac{b_x^2}{\Gamma_2}=
\frac{a^2}{\Gamma_1}-\frac{b^2}{\Gamma_2}.
\end{equation}
Let us introduce a new variable
\begin{equation}
y=\frac{a_x^2-a^2}{\Gamma_1}=\frac{b_x^2-b^2}{\Gamma_2}
\Leftrightarrow a_x=\sqrt{a^2+y\Gamma_1},\; b_x=\sqrt{b^2+y\Gamma_2}\;
\end{equation}
Now for two variables $\psi_x$ and $y$
we have two equations:
\begin{equation}\label{eq:System_for_y_psi}
\left\{
\begin{array}{l}
a_xb_x\cdot\left[\frac{\cos(2\psi_x)}{2}
\left(\Gamma_{1}+\Gamma_{2}\right)+\left(m_{1}-m_{2}\right)\sin(2\psi_x)\right]=
\\[3mm]\rule{20mm}{0mm}=
ab\cdot\left[\frac{\cos(2\psi)}{2}
\left(\Gamma_{1}+\Gamma_{2}\right)+\left(m_{1}-m_{2}\right)\sin(2\psi)\right]
-\\[3mm]\rule{30mm}{0mm}
-y\cdot\left[\left(m_{1}-m_{2}\right)^2+
\left(\frac{\Gamma_{1}+\Gamma_{2}}{2}\right)^2
\right],\\[10mm]
{a_xb_x\cdot\left[\left(m_{1}-m_{2}\right)\cos(2\psi_x)-
\frac{\sin(2\psi_x)}{2}
\left(\Gamma_{1}+\Gamma_{2}\right)\right]}
=\\[3mm]\rule{20mm}{0mm}
=
{ab\cdot\left[\left(m_{1}-m_{2}\right)\cos(2\psi)-
\frac{\sin(2\psi)}{2}
\left(\Gamma_{1}+\Gamma_{2}\right)\right]}.
\end{array}
\right.
\end{equation}

Linear equation for
 $\mathrm{tg}(2\psi_x)$ can be obtained dividing
 the first equation in (\ref{eq:System_for_y_psi}) on the second one:
\begin{equation}
\begin{array}{l}
\frac{\frac{\Gamma_{1}+\Gamma_{2}}{2}
+\left(m_{1}-m_{2}\right)\mathrm{tg}(2\psi_x)}
{\left(m_{1}-m_{2}\right)-
\frac{\mathrm{tg}(2\psi_x)}{2}
\left(\Gamma_{1}+\Gamma_{2}\right)}=
\\[3mm]\rule{10mm}{0mm}=
\frac{ab\cdot\left[\frac{\cos(2\psi)}{2}
\left(\Gamma_{1}+\Gamma_{2}\right)+\left(m_{1}-m_{2}\right)\sin(2\psi)\right]
-y\cdot\left[\left(m_{1}-m_{2}\right)^2+
\left(\frac{\Gamma_{1}+\Gamma_{2}}{2}\right)^2
\right]}
{ab\cdot\left[\left(m_{1}-m_{2}\right)\cos(2\psi)-
\frac{\sin(2\psi)}{2}
\left(\Gamma_{1}+\Gamma_{2}\right)\right]}
\end{array}
\end{equation}

The solution is amazingly simple:
\begin{equation}\label{eq:Solution_for_psix}
\mathrm{tg}2\psi_x=\frac{a\,b\sin(2\psi)+\left(m_2-m_1\right)y}
{a\,b\,\cos(2\psi)-\frac{y}{2}\left(\Gamma_1+\Gamma_2\right)}.
\end{equation}

Now we can check whether it is an actual solution of the
system~(\ref{eq:System_for_y_psi}).
Let us check the following values:
\begin{equation}
\begin{array}{l}
S_1=\frac{a_xb_x\cdot\left[\frac{\cos(2\psi_x)}{2}
\left(\Gamma_{1}+\Gamma_{2}\right)+\left(m_{1}-m_{2}\right)\sin(2\psi_x)\right]}
{ab\cdot\left[\frac{\cos(2\psi)}{2}
\left(\Gamma_{1}+\Gamma_{2}\right)+\left(m_{1}-m_{2}\right)\sin(2\psi)\right]
-y\cdot\left[\left(m_{1}-m_{2}\right)^2+
\left(\frac{\Gamma_{1}+\Gamma_{2}}{2}\right)^2
\right]},\\[7mm]
S_2=\frac{a_xb_x\cdot\left[\left(m_{1}-m_{2}\right)\cos(2\psi_x)-
\frac{\sin(2\psi_x)}{2}
\left(\Gamma_{1}+\Gamma_{2}\right)\right]}
{ab\cdot\left[\left(m_{1}-m_{2}\right)\cos(2\psi)-
\frac{\sin(2\psi)}{2}
\left(\Gamma_{1}+\Gamma_{2}\right)\right]}.
\end{array}
\end{equation}
For an actual solution there should be $S_1=1,\;\; S_2=1$.
After substitution $\psi_x$ we obtain:
\begin{equation}
S_1=S_2=\frac{a_xb_x\cos(2\psi_x)}{ab\cos(2\psi)-\frac{y}{2}\cdot\left(\Gamma_1+\Gamma_2
\right)}
\end{equation}

From $S_{1,2}=1$:
\begin{equation}
\frac{\cos(2\psi_x)}{ab\,\cos(2\psi)-\frac{y}{2}\cdot\left(\Gamma_1+\Gamma_2\right)}=
\frac{1}{a_xb_x}=\frac{1}{\sqrt{\left(a^2+y\Gamma_1\right)\left(b^2+y\Gamma_2\right)}},
\end{equation}
and from expression~(\ref{eq:Solution_for_psix}) we get:
\begin{equation}
\begin{array}{l}
\frac{\cos(2\psi_x)}{ab\cos(2\psi)-\frac{y}{2}\cdot\left(\Gamma_1+\Gamma_2\right)}=
\\ =
\frac{1}{\sqrt{
a^2b^2+y^2\cdot\left[(m_1-m_2)^2+\left(\frac{\Gamma_1+\Gamma_2}{2}\right)^2\right]+
2a\,b\,y\cdot\left[(m_2-m_1)\sin(2\psi)-\frac{\Gamma_1+\Gamma_2}{2}\cos(2\psi)\right]
}}
\end{array}
\end{equation}

So the equation for $y$:
\begin{equation}
\begin{array}{l}
y\cdot\left\{y\cdot
\left[(m_1-m_2)^2+\left(\frac{\Gamma_1+\Gamma_2}{2}\right)^2
-\Gamma_1\Gamma_2\right]+
\right. \\[4mm]\left.\rule{3mm}{0mm}+
2a\,b\cdot\left[(m_2-m_1)\sin(2\psi)-\frac{\Gamma_1+\Gamma_2}{2}\cos(2\psi)
\right]-a^2\Gamma_2-b^2\Gamma_1\right\}=0
\end{array}
\end{equation}

This equation has two solutions for $y$:

\framebox{$y=0$} \ \\
$a_x=a,\; b_x=b,\; \mathrm{tg}(2\psi_x)=\mathrm{tg}(2\psi),\;
S_1=S_2=\frac{\cos(2\psi_x)}{\cos(2\psi)}$.\\
It is obvious that the values $\sin(2\psi_x)$ and $\cos(2\psi_x)$
must match exactly with  $\sin(2\psi)$ and $\cos(2\psi)$, correspondingly.


\framebox{$y=\frac{a^2\Gamma_2+b^2\Gamma_1+2ab\cdot\left[\frac{\cos(2\psi)}{2}
\left(\Gamma_{1}+\Gamma_{2}\right)+\left(m_{1}-m_{2}\right)\sin(2\psi)\right]}
{\left(m_{1}-m_{2}\right)^2+\left(\frac{\Gamma_{1}-\Gamma_{2}}{2}\right)^2}
$} \ \\[4mm]

\begin{equation}
a_x=\sqrt{a^2+y\Gamma_1},\;\; b_x=\sqrt{b^2+y\Gamma_2},
\end{equation}
\begin{equation}
\cos(2\psi_x)=\frac{ab\cos(2\psi)-\frac{y}{2}\left(\Gamma_1+\Gamma_2\right)}
{a_xb_x},
\end{equation}
\begin{equation}
\sin(2\psi_x)=\frac{ab\sin(2\psi)+(m_2-m_1)y}{a_xb_x}.
\end{equation}
If one formally substitutes this solution, then  two cross section curves
become identical.
However in order to this solution be admittable some conditions
must be satisfied:
\begin{enumerate}
\item
$y\geq -\frac{a^2}{\Gamma_1}\Longleftrightarrow a_x^2\geq 0$,
\item
$y\geq -\frac{b^2}{\Gamma_2}\Longleftrightarrow b_x^2\geq 0$,
\item
$\left|\cos(2\psi_x)\right| \leq 1,$
\item
$\left|\sin(2\psi_x)\right| \leq 1.$
\end{enumerate}
Let us check
\begin{equation}
\begin{array}{l}
a_x^2=a^2+\Gamma_1y=
\\[4mm]\rule{5mm}{0mm}=
\frac{
b^2\Gamma_1^2+a^2\cdot\left[
\left(m_{1}-m_{2}\right)^2+\left(\frac{\Gamma_{1}+\Gamma_{2}}{2}\right)^2\right]
+2ab\Gamma_1\cdot\left[\frac{\cos(2\psi)}{2}
\left(\Gamma_{1}+\Gamma_{2}\right)+\left(m_{1}-m_{2}\right)\sin(2\psi)\right]
}
{\left(m_{1}-m_{2}\right)^2+\left(\frac{\Gamma_{1}-\Gamma_{2}}{2}\right)^2}=
\\[4mm]\rule{5mm}{0mm}= \frac{
\left[a\sqrt{\left(m_{1}-m_{2}\right)^2+\left(\frac{\Gamma_{1}+\Gamma_{2}}{2}\right)^2}
+b\Gamma_1\sin\left(2\psi+\mathrm{arctg}\frac{\Gamma_1+\Gamma_2}{2(m_1-m_2)}\right)
\right]^2}
{\left(m_{1}-m_{2}\right)^2+\left(\frac{\Gamma_{1}-\Gamma_{2}}{2}\right)^2}
+\\[4mm] \rule{20mm}{0mm}
+\frac{
b^2\Gamma_1^2\cos^2
\left(2\psi+\mathrm{arctg}\frac{\Gamma_1+\Gamma_2}{2(m_1-m_2)}\right)
}
{\left(m_{1}-m_{2}\right)^2+\left(\frac{\Gamma_{1}-\Gamma_{2}}{2}\right)^2}
\geq 0.
\end{array}
\end{equation}

Similarly the condition $b_x^2\geq 0$ is checked.
The last two conditions are easily confirmed by the check that
derived by their own formulae  $\sin(2\psi_x)$
and
$\cos(2\psi_x)$ satisfy the Pythagorean theorem
$\cos^2(2\psi_x)+\sin^2(2\psi_x)=1$.

So we have found non-trivial solution that means: any pair of resonances
can be replaced by another pair of resonances with the same
masses and widths but different amplitudes and phases so that
the resonance cross section does not change.

Using these formulae one can get the solution for the case of
interference of Breit-Wigner amplitude with complex constant,
substituting

$$
b\to c\cdot i\frac{\Gamma_2}{2}
$$
and setting $\Gamma_2\to\infty$.
Corresponding cross section reads
\begin{equation}\label{eq:Section_wave}
\sigma(E)=\left|\frac{ae^{i\psi}}{E-m_1+i\frac{\Gamma_1}{2}}+
ce^{-i\psi}\right|^2.
\end{equation}

Finally we get
\begin{equation}
\begin{array}{l}
y=\!\!\!%
\lim\limits_{\Gamma_2\to\infty}
\frac{a^2\Gamma_2+\left(c\cdot i\frac{\Gamma_2}{2}\right)^2
\Gamma_1+a\cdot c \cdot i\cdot \Gamma_2\cdot\left[\frac{\cos(2\psi)}{2}
\left(\Gamma_{1}+\Gamma_{2}\right)+\left(m_{1}-m_{2}\right)\sin(2\psi)\right]}
{\left(m_{1}-m_{2}\right)^2+\left(\frac{\Gamma_{1}-\Gamma_{2}}{2}\right)^2}=
\\[4mm] \rule{60mm}{0mm}=
2iac\cos 2\psi-c^2\Gamma_1,\\[3mm]
a_x=\sqrt{a^2+2iac\Gamma_1\cos 2\psi -c^2\Gamma_1^2},\;
c_x=c,\\[3mm]
\mathrm{tg} 2\psi_x=\frac{a\sin 2\psi}{a\cos 2\psi + i\frac{y}{c}}.
\end{array}
\end{equation}
Here the variable
 $y$, which was real in previous consideration
became complex as well as variable  $a_x$.
If reassemble imaginary and real parts of these variables
or solve this problem from the beginning with cross
section~(\ref{eq:Section_wave}),
then one gets
\begin{equation}\!\!%
\begin{array}{l}
a_x=\sqrt{a^2+2ac\Gamma_1\sin 2\psi+c^2\Gamma_1^2}=
\sqrt{\left(c\Gamma_1+a\sin 2\psi\right)^2+a^2\cos^2 2\psi},\;
\\[3mm]
\sin 2\psi_x=-\frac{a\sin 2\psi +c\Gamma_1}{a_x},\;
\cos 2\psi_x=\frac{a\cos 2\psi}{a_x}.
\end{array}
\end{equation}

Let us look at the numerical example of $\omega$ and $\phi$
mesons interference in the channel $e^+e^-\to\pi^+\pi^-\pi^0$.
Approximate values of resonance parameters:
$m_1=m_\omega=782.6$, $\Gamma_\omega=8.4$, $a=1$,
$m_2=m_\phi=1019.4$, $\Gamma_\phi=4.46$, $b=0.1$,
$\psi=-\frac{155^\circ}{2}=-77.5^\circ$.

Substituting to the formulae one gets
\begin{equation}
y=0.00042; \;\; a_x=1.0018;\;\; b_x=0.109; \;\; \psi_x=74.4^\circ.
\end{equation}

So we get quite different phase while amplitudes changed just a little bit.

At this point the most urgent question is: whether this
ambiguity is a unique property of just this simple resonance
description, or in a more sophisticated and realistic parameterization
of resonance cross section such resonance phase ambiguity
will take place?
The matter is that for actual experimental data processing
the much more complicated resonance cross section formulae are used.

\section{Relativistic Breit-Wigner resonance amplitude
}

A little more complicated variant of Breit-Wigner formula
(relativistic) is:
\begin{equation}
\sigma(E)=\left|\frac{A}{s-m_1^2+i{\Gamma_1}m_1}+
\frac{B}{s-m_2^2+i{\Gamma_2}m_2}\right|^2,
\end{equation}
where  $s=E^2$,
evidently has the same property, just some redefinition
of variables is necessary.

It is also evident that additional general factor, even
strongly dependent on energy, does not change the solution.

However in the most accurate variant of this formula
instead of constants $\Gamma_1$, $\Gamma_2$
there are used more complicated expressions containing
phase space of final states, transition to which are
probable for these resonances.

\section{Energy dependence of resonance width}

For multiparticle final states, such as  $\pi^+\pi^-\pi^0$,
there are no simple formulae for the final state phase space,
so for our exercise let us choose a simple model formula:
\begin{equation}
\Gamma_i\to
\Gamma_i\cdot\left(\frac{s-{4\mu^2}}{{m_i^2}-{4\mu^2}}
\right)^{\frac{3}{4}},
\end{equation}
where the effective mass $\mu$ defines the reaction threshold,
so at $s<4\mu^2$  cross section becomes equal zero.
New cross section can be rewritten as follows
\begin{equation}
\begin{array}{l}
\sigma(s)=\frac{\left(s-{4\mu^2}\right)^{\frac{3}{2}}}{s^2}
\times
\\[4mm] \rule{3mm}{0mm}\times
\left|
\frac{A}{s-m_1^2+i{\Gamma_1}m_1\cdot
\left(\frac{s-{4\mu^2}}{{m_1^2}-{4\mu^2}}
\right)^{\frac{3}{4}}}+
\frac{B}{s-m_2^2+i{\Gamma_2}m_2\cdot
\left(\frac{s-{4\mu^2}}{{m_2^2}-{4\mu^2}}
\right)^{\frac{3}{4}}}\right|^2
\end{array}
\end{equation}

For this function it is hard to make a Fourrier transform in order
to apply the same trick we have done for the approximate
resonance curve. Substitute
\begin{equation}
\begin{array}{l}
s=4\mu^2+\rho^4,\;\;
\rho=\left(s-{4\mu^2}\right)^{\frac{1}{4}},\;\; \\[4mm]
\rho_1=\left({m_1^2}-{4\mu^2}\right)^{\frac{1}{4}},\;\;
\rho_2=\left({m_2^2}-{4\mu^2}\right)^{\frac{1}{4}}.
\end{array}
\end{equation}

Cross section dependence on the new variable $\rho$
is the following
\begin{equation}
\begin{array}{l}
\sigma(\rho)=\frac{\rho^6}{\left(\rho^4+{4\mu^2}\right)^2}
\times
\\[4mm] \rule{3mm}{0mm}\times
\left|
\frac{A}{\rho^4+{4\mu^2}-m_1^2+i{\Gamma_1}m_1\cdot
\left(\frac{\rho}{\rho_1}
\right)^{3}}+
\frac{B}{\rho^4+{4\mu^2}-m_2^2+i{\Gamma_2}m_2\cdot
\left(\frac{\rho}{\rho_2}
\right)^{3}}\right|^2
\end{array}
\end{equation}

The task of Fourrier transform of this function is
already not so hard if we can find all its irregular
points.
Just let us simplify the function first
so as the general factor does not change the problem solution.

Thus investigated function of $\rho$ is
\begin{equation}
\begin{array}{l}
f(\rho)=
\left|
\frac{A}{\rho^4+{4\mu^2}-m_1^2+i{\Gamma_1}m_1\cdot
\left(\frac{\rho}{\rho_1}
\right)^{3}}+
\frac{B}{\rho^4+{4\mu^2}-m_2^2+i{\Gamma_2}m_2\cdot
\left(\frac{\rho}{\rho_2}
\right)^{3}}\right|^2
\end{array}
\end{equation}

16 irregular points are determined by the equations
\begin{equation}
\rho^4+{4\mu^2}-m_j^2\pm i{\Gamma_j}m_j\cdot
\left(\frac{\rho}{\rho_j}
\right)^{3}=0,\;\; j=1,2
\end{equation}

Rewrite this equation in the form
\begin{equation}\label{eq:Fourth_power}
\rho^4+R_1\rho^3+R_2=0,
\end{equation}
where $R_1=\pm i\frac{{\Gamma_j}m_j}{\rho_j^{3}}$,
$R_2={4\mu^2}-m_j^2=-\rho_j^4$.

If we solve this equation in a standard way, then
the solution can be excessive complicated.
Let us try to apply the idea of Ferrari solution
directly to the equation~(\ref{eq:Fourth_power}).
\begin{equation}
\rho^4+R_1\rho^3+R_2=\left(\rho^2+\lambda_1\rho+\lambda_2\right)\cdot
\left(\rho^2+\lambda_3\rho+\lambda_4\right)
\end{equation}
This equality is valid for any $\rho$ if
\begin{equation}
\left\{
\begin{array}{l}
\lambda_1+\lambda_3=R_1\\
\lambda_2+\lambda_4+\lambda_1\lambda_3=0\\
\lambda_2\lambda_3+\lambda_1\lambda_4=0\\
\lambda_2\lambda_4=R_2
\end{array}\right.
\end{equation}

Exclude $\lambda_1$, $\lambda_3$:
\begin{equation}
\lambda_1=\frac{\lambda_2}{\lambda_2-\lambda_4}R_1,\;\;\;
\lambda_3=\frac{\lambda_4}{\lambda_4-\lambda_2}R_1.
\end{equation}

Now for the variables $\lambda_2$ and $\lambda_4$
we have two equations:
\begin{equation}
\left\{
\begin{array}{l}
\left(\lambda_2+\lambda_4\right)\cdot\left(\lambda_2^2+\lambda_4^2-2R_2\right)-R_1^2R_2=0,\\
\lambda_2\lambda_4=R_2.
\end{array}\right.
\end{equation}

Substituting $\lambda_2+\lambda_4=y$, obtain
\begin{equation}
\begin{array}{l}
\lambda_1=\frac{y+\sqrt{y^2+4\rho_j^4}}{2\sqrt{y^2+4\rho_j^4}}R_1,\\
\lambda_2=\frac{1}{2}\left(y+\sqrt{y^2+4\rho_j^4}\right),\\
\lambda_3=-\frac{y-\sqrt{y^2+4\rho_j^4}}{2\sqrt{y^2+4\rho_j^4}}R_1,\\
\lambda_4=\frac{1}{2}\left(y-\sqrt{y^2+4\rho_j^4}\right),
\end{array}
\end{equation}
and $y$ must satisfy the equation
\begin{equation}
y^3-4R_2y-R_1^2R_2=0.
\end{equation}

Applying Cardano solution~\cite{KORN}:
\begin{equation}
y=\alpha+\beta \Longrightarrow \alpha^3+\beta^3 +\left(\alpha+\beta\right)\cdot
\left(3\alpha\beta-4R_2\right)-R_1^2R_2=0.
\end{equation}
Setting $\beta= \frac{4R_2}{3\alpha}$, we get
\begin{equation}
\alpha^6-R_1^2R_2\alpha^3+\left(\frac{4R_2}{3}\right)^3=0,
\end{equation}
\begin{equation}
\begin{array}{l}
\alpha=\sqrt[3]{\frac{1}{2}R_1^2R_2 +\sqrt{\left(\frac{R_1^2R_2}{2}\right)^2-
\left(\frac{4R_2}{3}\right)^3}},\\[4mm]
\beta=\sqrt[3]{\frac{1}{2}R_1^2R_2 -\sqrt{\left(\frac{R_1^2R_2}{2}\right)^2-
\left(\frac{4R_2}{3}\right)^3}}.
\end{array}
\end{equation}

Substituting $R_1$, $R_2$, we get
\begin{equation}
\begin{array}{l}
\alpha=\sqrt[3]{\frac{\Gamma_j^2m_j^2}{2\rho_j^2}+
\sqrt{\left(\frac{\Gamma_j^2m_j^2}{2\rho_j^2}\right)^2
+\left(\frac{4\rho_j^4}{3}\right)^3}}=
\\[4mm]\rule{4mm}{0mm}=
\rho_j^2\sqrt[3]{\frac{\Gamma_j^2m_j^2}{2\rho_j^8}+
\sqrt{\left(\frac{\Gamma_j^2m_j^2}{2\rho_j^8}\right)^2
+\left(\frac{4}{3}\right)^3}}.
\end{array}
\end{equation}

It can be easily confirmed that $\alpha>0$, $\beta<0$, $y>0$,
$\lambda_2>0$,  $\lambda_4<0$ are the real numbers,
$\lambda_1$, $\lambda_3$ are complex numbers.

Two roots are determined by the equation
\begin{equation}
\rho^2+\lambda_1\rho+\lambda_2=0,
\end{equation}
and two more roots by the equation
\begin{equation}
\rho^2+\lambda_3\rho+\lambda_4=0.
\end{equation}
None of these roots can be a real number that is evident from the
initial form of quatric equation: $\rho=0$ is not root,
and for any real  $\rho\neq 0$ polynomial has non-zero imaginary
part for non-zero width $\Gamma_j>0$ and mass $m_j>0$.
In order to calculate integrals with infinite limits by residue method
we are interested to know the sign of imaginary part of roots.
Introduce notation for all roots.

\begin{equation}
\begin{array}{l}
\lambda_{11}=i\frac{y_1+\sqrt{y_1^2+4\rho_1^4}}{2\sqrt{y_1^2+4\rho_1^4}}
\frac{\Gamma_1m_1}{\rho_1^3},\\
\lambda_{21}=\frac{1}{2}\left(y_1+\sqrt{y_1^2+4\rho_1^4}\right),\\
\lambda_{31}=-i\frac{y_1-\sqrt{y_1^2+4\rho_1^4}}{2\sqrt{y_1^2+4\rho_1^4}}
\frac{\Gamma_1m_1}{\rho_1^3},\\
\lambda_{41}=\frac{1}{2}\left(y_1-\sqrt{y_1^2+4\rho_1^4}\right),\\
y_1=\rho_1^2\cdot\left(\sqrt[3]{\frac{\Gamma_1^2m_1^2}{2\rho_1^8}+
\sqrt{\left(\frac{\Gamma_1^2m_1^2}{2\rho_1^8}\right)^2
+\left(\frac{4}{3}\right)^3}}+
\right. \\[3mm]\rule{15mm}{0mm}+\left.
\sqrt[3]{\frac{\Gamma_1^2m_1^2}{2\rho_1^8}-
\sqrt{\left(\frac{\Gamma_1^2m_1^2}{2\rho_1^8}\right)^2
+\left(\frac{4}{3}\right)^3}}
\right),\\
z_{11}=-\frac{\lambda_{11}}{2}+\sqrt{\frac{\lambda_{11}^2}{4}-\lambda_{21}},\\
z_{21}=-\frac{\lambda_{11}}{2}-\sqrt{\frac{\lambda_{11}^2}{4}-\lambda_{21}},\\
z_{31}=-\frac{\lambda_{31}}{2}+\sqrt{\frac{\lambda_{31}^2}{4}-\lambda_{41}},\\
z_{41}=-\frac{\lambda_{31}}{2}-\sqrt{\frac{\lambda_{31}^2}{4}-\lambda_{41}}
\end{array}
\end{equation}

One can see that $\Re(z_{11})=\Re(z_{21})=0$, $\Im(z_{11})>0$, $\Im(z_{21})<0$,
$\Im(z_{31})<0$, $\Im(z_{41})<0$.

Four roots correspond better to the equation with index $j=2$:
\begin{equation}
\begin{array}{l}
\lambda_{12}=i\frac{y_2+\sqrt{y_2^2+4\rho_2^4}}{2\sqrt{y_2^2+4\rho_2^4}}
\frac{\Gamma_2m_2}{\rho_2^3},\\
\lambda_{22}=\frac{1}{2}\left(y_2+\sqrt{y_2^2+4\rho_2^4}\right),\\
\lambda_{32}=-i\frac{y_2-\sqrt{y_2^2+4\rho_2^4}}{2\sqrt{y_2^2+4\rho_2^4}}
\frac{\Gamma_2m_2}{\rho_2^3},\\
\lambda_{42}=\frac{1}{2}\left(y_2-\sqrt{y_2^2+4\rho_2^4}\right),\\
y_2=\rho_2^2\cdot\left(\sqrt[3]{\frac{\Gamma_2^2m_2^2}{2\rho_2^8}+
\sqrt{\left(\frac{\Gamma_2^2m_2^2}{2\rho_2^8}\right)^2
+\left(\frac{4}{3}\right)^3}}+
\right. \\[3mm]\rule{15mm}{0mm}+\left.
\sqrt[3]{\frac{\Gamma_2^2m_2^2}{2\rho_2^8}-
\sqrt{\left(\frac{\Gamma_2^2m_2^2}{2\rho_2^8}\right)^2
+\left(\frac{4}{3}\right)^3}}
\right),\\
z_{12}=-\frac{\lambda_{12}}{2}+\sqrt{\frac{\lambda_{12}^2}{4}-\lambda_{22}},\\
z_{22}=-\frac{\lambda_{12}}{2}-\sqrt{\frac{\lambda_{12}^2}{4}-\lambda_{22}},\\
z_{32}=-\frac{\lambda_{32}}{2}+\sqrt{\frac{\lambda_{32}^2}{4}-\lambda_{42}},\\
z_{42}=-\frac{\lambda_{32}}{2}-\sqrt{\frac{\lambda_{32}^2}{4}-\lambda_{42}}
\end{array}
\end{equation}

Here $\Re(z_{12})=\Re(z_{22})=0$, $\Im(z_{12})>0$, $\Im(z_{22})<0$,
$\Im(z_{32})<0$, $\Im(z_{42})<0$.

Now the function $f(\rho)$ can be presented in the form
\begin{equation}
\begin{array}{l}
f(\rho)=\left(\frac{A}{\left(\rho-z_{11}\right)\left(\rho-z_{21}\right)
\left(\rho-z_{31}\right)\left(\rho-z_{41}\right)}+
\frac{B}{\left(\rho-z_{12}\right)\left(\rho-z_{22}\right)
\left(\rho-z_{32}\right)\left(\rho-z_{42}\right)}\right)\times
\\[4mm]\rule{4mm}{0mm}\times
\left(\frac{A^*}{\left(\rho-z_{11}^*\right)\left(\rho-z_{21}^*\right)
\left(\rho-z_{31}^*\right)\left(\rho-z_{41}^*\right)}+
\frac{B^*}{\left(\rho-z_{12}^*\right)\left(\rho-z_{22}^*\right)
\left(\rho-z_{32}^*\right)\left(\rho-z_{42}^*\right)}\right),
\end{array}
\end{equation}
where symbol $^{*}$ designates complex conjugate number.
Fourrier image  $F(t)=\int\limits_{-\infty}^{+\infty}f(\rho)e^{it\rho}\df\rho$
of the real function has a property that $F(-t)=F^*(t)$,
so it is enough to calculate Fourrier transform only for positive value
of $t$, that is determined by the sum of residues on the irregular points
above the abscissa axis, that is $z_{11}$, $z_{21}^*$, $z_{31}^*$, $z_{41}^*$,
$z_{12}$,  $z_{22}^*$, $z_{32}^*$, $z_{42}^*$.
\begin{equation}\label{eq:FourrierImage}
\begin{array}{l}
F(t)=2\pi i\times \\[5mm] \times\left\{
\frac{Ae^{itz_{11}}}{\left(z_{11}-z_{21}\right)
\left(z_{11}-z_{31}\right)\left(z_{11}-z_{41}\right)}\cdot
\left(\frac{A^*}{2z_{11}\left(z_{11}-z_{21}^*\right)
\left(z_{11}-z_{31}^*\right)\left(z_{11}-z_{41}^*\right)}+\right.\right. \\[3mm]
\left.\rule{20mm}{0mm}+
\frac{B^*}{\left(z_{11}-z_{12}^*\right)\left(z_{11}-z_{22}^*\right)
\left(z_{11}-z_{32}^*\right)\left(z_{11}-z_{42}^*\right)}\right)+\\[3mm]
+\frac{A^*e^{itz_{21}^*}}{\left(z_{21}^*-z_{11}^*\right)
\left(z_{21}^*-z_{31}^*\right)\left(z_{21}^*-z_{41}^*\right)}\cdot
\left(\frac{A}{\left(z_{21}^*-z_{11}\right)2z_{21}^*
\left(z_{21}^*-z_{31}\right)\left(z_{21}^*-z_{41}\right)}+\right.\\[3mm]
\left.\rule{20mm}{0mm}+
\frac{B}{\left(z_{21}^*-z_{12}\right)\left(z_{21}^*-z_{22}\right)
\left(z_{21}^*-z_{32}\right)\left(z_{21}^*-z_{42}\right)}\right)+\\[3mm]
+\frac{A^*e^{itz_{31}^*}}{\left(z_{31}^*-z_{11}^*\right)\left(z_{31}^*-z_{21}^*\right)
\left(z_{31}^*-z_{41}^*\right)}\cdot
\left(\frac{A}{\left(z_{31}^*-z_{11}\right)\left(z_{31}^*-z_{21}\right)
\left(z_{31}^*-z_{31}\right)\left(z_{31}^*-z_{41}\right)}+\right.\\[3mm]
\rule{20mm}{0mm}+\left.
\frac{B}{\left(z_{31}^*-z_{12}\right)\left(z_{31}^*-z_{22}\right)
\left(z_{31}^*-z_{32}\right)\left(z_{31}^*-z_{42}\right)}\right)+\\[3mm]
+\frac{A^*e^{itz_{41}^*}}{\left(z_{41}^*-z_{11}^*\right)\left(z_{41}^*-z_{21}^*\right)
\left(z_{41}^*-z_{31}^*\right)}\cdot
\left(\frac{A}{\left(z_{41}^*-z_{11}\right)\left(z_{41}^*-z_{21}\right)
\left(z_{41}^*-z_{31}\right)\left(z_{41}^*-z_{41}\right)}+\right.\\[3mm]
\left.\rule{20mm}{0mm}+
\frac{B}{\left(z_{41}^*-z_{12}\right)\left(z_{41}^*-z_{22}\right)
\left(z_{41}^*-z_{32}\right)\left(z_{41}^*-z_{42}\right)}\right)+\\[3mm]
+\frac{Be^{itz_{12}}}{\left(z_{12}-z_{22}\right)
\left(z_{12}-z_{32}\right)\left(z_{12}-z_{42}\right)}\cdot
\left(\frac{A^*}{\left(z_{12}-z_{11}^*\right)\left(z_{12}-z_{21}^*\right)
\left(z_{12}-z_{31}^*\right)\left(z_{12}-z_{41}^*\right)}+\right. \\[3mm]
\left.\rule{20mm}{0mm}+
\frac{B^*}{2z_{12}\left(z_{12}-z_{22}^*\right)
\left(z_{12}-z_{32}^*\right)\left(z_{12}-z_{42}^*\right)}\right)+\\[3mm]
+\frac{B^*e^{itz_{22}^*}}{\left(z_{22}^*-z_{12}^*\right)
\left(z_{22}^*-z_{32}^*\right)\left(z_{22}^*-z_{42}^*\right)}\cdot
\left(\frac{A}{\left(z_{22}^*-z_{11}\right)\left(z_{22}^*-z_{21}\right)
\left(z_{22}^*-z_{31}\right)\left(z_{22}^*-z_{41}\right)}+\right.\\[3mm]
\left.\rule{20mm}{0mm}+
\frac{B}{\left(z_{22}^*-z_{12}\right)2z_{22}^*
\left(z_{22}^*-z_{32}\right)\left(z_{22}^*-z_{42}\right)}\right)+\\[3mm]
+\frac{B^*e^{itz_{32}^*}}{\left(z_{32}^*-z_{12}^*\right)\left(z_{32}^*-z_{22}^*\right)
\left(z_{32}^*-z_{42}^*\right)}\cdot
\left(\frac{A}{\left(z_{32}^*-z_{11}\right)\left(z_{32}^*-z_{21}\right)
\left(z_{32}^*-z_{31}\right)\left(z_{32}^*-z_{41}\right)}+\right.\\[3mm]
\rule{20mm}{0mm}+\left.
\frac{B}{\left(z_{32}^*-z_{12}\right)\left(z_{32}^*-z_{22}\right)
\left(z_{32}^*-z_{32}\right)\left(z_{32}^*-z_{42}\right)}\right)+\\[3mm]
+\frac{B^*e^{itz_{42}^*}}{\left(z_{42}^*-z_{12}^*\right)\left(z_{42}^*-z_{22}^*\right)
\left(z_{42}^*-z_{32}^*\right)}\cdot
\left(\frac{A}{\left(z_{42}^*-z_{11}\right)\left(z_{42}^*-z_{21}\right)
\left(z_{42}^*-z_{31}\right)\left(z_{42}^*-z_{41}\right)}+\right.\\[3mm]
\left.\rule{20mm}{0mm}+
\left.
\frac{B}{\left(z_{42}^*-z_{12}\right)\left(z_{42}^*-z_{22}\right)
\left(z_{42}^*-z_{32}\right)\left(z_{42}^*-z_{42}\right)}\right)
\right\}
\end{array}
\end{equation}

Because the coordinates of irregular points determine
the functional dependence of Fourrier image on the parameter $t$,
so they must be unchanged for any
solution. This condition produces the equations:
\begin{equation}
\rho^4+4\mu_x^2-m_{jx}^2\pm i\Gamma_{jx}m_{jx}\left(\frac{\rho}{\rho_{jx}}\right)^3\equiv
\rho^4+4\mu^2-m_{j}^2\pm i\Gamma_{j}m_{j}\left(\frac{\rho}{\rho_{j}}\right)^3
\end{equation}
for $j=1,2$.
Thus we get the system of equations:
\begin{equation}
\left\{\begin{array}{l}
4\mu_x^2-m_{1x}^2=4\mu^2-m_1^2,\\
\frac{\Gamma_{1x}m_{1x}}{\rho_{1x}^3}=\frac{\Gamma_{1}m_{1}}{\rho_{1}^3},\\
4\mu_x^2-m_{2x}^2=4\mu^2-m_2^2,\\
\frac{\Gamma_{2x}m_{2x}}{\rho_{2x}^3}=\frac{\Gamma_{2}m_{2}}{\rho_{2}^3},\\
\end{array}\right.
\end{equation}
where
$\rho_1=\left(m_1^2-4\mu^2\right)^{1/4},\rho_2=\left(m_2^2-4\mu^2\right)^{1/4}$.
For five variables $\mu_x,\Gamma_{1x},\Gamma_{2x},m_{1x},m_{2x}$
we got the system of the four equations.
However the factor  $\rho^3$ in the original function
determined the threshold behaviour, and from $\rho=\left(s-4\mu^2\right)^{1/4}$
is evident that
$\mu$ must not change, so
\begin{equation}
\mu_x=\mu
\end{equation}
and we have four equations for the four variables.
Taking into account that $m_{jx}>0$,
we derive $m_{jx}=m_j$ and can
immediately conclude that $\rho_{jx}=\rho_j$,
and  $\Gamma_{jx}=\Gamma_j$.

Let us try to simplify the denominators in the
formula~(\ref{eq:FourrierImage}).
Apparently
\begin{equation}
\begin{array}{l}
2z_{11}\left(z_{11}-z_{21}^*\right)
\left(z_{11}-z_{31}^*\right)\left(z_{11}-z_{41}^*\right)=
\\[4mm] \rule{3mm}{0mm}=
z_{11}^4+4\mu^2-m_1^2-i\Gamma_1m_1\left(\frac{z_{11}}{\rho_1}\right)^3=
\\[4mm]\rule{3mm}{0mm}=
z_{11}^4+4\mu^2-m_1^2-i\Gamma_1m_1\left(\frac{z_{11}}{\rho_1}\right)^3-
\\[3mm]\rule{15mm}{0mm}-
\left[z_{11}^4+4\mu^2-m_1^2+i\Gamma_1m_1\left(\frac{z_{11}}{\rho_1}\right)^3
\right]=-2i\Gamma_1m_1\left(\frac{z_{11}}{\rho_1}\right)^3
\end{array}
\end{equation}

Similarly
\begin{equation}
\left(z_{21}^*-z_{11}\right)2z_{21}^*
\left(z_{21}^*-z_{31}\right)\left(z_{21}^*-z_{41}\right)=
2i\Gamma_1m_1\left(\frac{z_{21}^*}{\rho_1}\right)^3
\end{equation}
\begin{equation}
\left(z_{31}^*-z_{11}\right)\left(z_{31}^*-z_{21}\right)
\left(z_{31}^*-z_{31}\right)\left(z_{31}^*-z_{41}\right)=
2i\Gamma_1m_1\left(\frac{z_{31}^*}{\rho_1}\right)^3
\end{equation}
\begin{equation}
\left(z_{41}^*-z_{11}\right)\left(z_{41}^*-z_{21}\right)
\left(z_{41}^*-z_{31}\right)\left(z_{41}^*-z_{41}\right)=
2i\Gamma_1m_1\left(\frac{z_{41}^*}{\rho_1}\right)^3
\end{equation}
\begin{equation}
2z_{12}\left(z_{12}-z_{22}^*\right)
\left(z_{12}-z_{32}^*\right)\left(z_{12}-z_{42}^*\right)=
-2i\Gamma_2m_2\left(\frac{z_{12}}{\rho_2}\right)^3
\end{equation}
\begin{equation}
\left(z_{22}^*-z_{12}\right)2z_{22}^*
\left(z_{22}^*-z_{32}\right)\left(z_{22}^*-z_{42}\right)=
2i\Gamma_2m_2\left(\frac{z_{22}^*}{\rho_2}\right)^3
\end{equation}
\begin{equation}
\left(z_{32}^*-z_{12}\right)\left(z_{32}^*-z_{22}\right)
\left(z_{32}^*-z_{32}\right)\left(z_{32}^*-z_{42}\right)=
2i\Gamma_2m_2\left(\frac{z_{32}^*}{\rho_2}\right)^3
\end{equation}
\begin{equation}
\left(z_{42}^*-z_{12}\right)\left(z_{42}^*-z_{22}\right)
\left(z_{42}^*-z_{32}\right)\left(z_{42}^*-z_{42}\right)=
2i\Gamma_2m_2\left(\frac{z_{42}^*}{\rho_2}\right)^3
\end{equation}

Eight other denominators are not so compact:
\begin{equation}
\begin{array}{l}
\left(z_{11}-z_{12}^*\right)\left(z_{11}-z_{22}^*\right)
\left(z_{11}-z_{32}^*\right)\left(z_{11}-z_{42}^*\right)=
\\[3mm]\rule{20mm}{0mm}=
-i\Gamma_2m_2\left(\frac{z_{11}}{\rho_2}\right)^3
-i\Gamma_1m_1\left(\frac{z_{11}}{\rho_1}\right)^3
-m_2^2+m_1^2
\end{array}
\end{equation}
\begin{equation}
\begin{array}{l}
\left(z_{21}^*-z_{12}\right)\left(z_{21}^*-z_{22}\right)
\left(z_{21}^*-z_{32}\right)\left(z_{21}^*-z_{42}\right)=
\\[3mm]\rule{20mm}{0mm}=
i\Gamma_2m_2\left(\frac{z_{21}^*}{\rho_2}\right)^3
+i\Gamma_1m_1\left(\frac{z_{21}^*}{\rho_1}\right)^3
-m_2^2+m_1^2
\end{array}
\end{equation}
\begin{equation}
\begin{array}{l}
\left(z_{31}^*-z_{12}\right)\left(z_{31}^*-z_{22}\right)
\left(z_{31}^*-z_{32}\right)\left(z_{31}^*-z_{42}\right)=
\\[3mm]\rule{20mm}{0mm}=
i\Gamma_2m_2\left(\frac{z_{31}^*}{\rho_2}\right)^3
+i\Gamma_1m_1\left(\frac{z_{31}^*}{\rho_1}\right)^3
-m_2^2+m_1^2
\end{array}
\end{equation}
\begin{equation}
\begin{array}{l}
\left(z_{41}^*-z_{12}\right)\left(z_{41}^*-z_{22}\right)
\left(z_{41}^*-z_{32}\right)\left(z_{41}^*-z_{42}\right)=
\\[3mm]\rule{20mm}{0mm}=
i\Gamma_2m_2\left(\frac{z_{41}^*}{\rho_2}\right)^3
+i\Gamma_1m_1\left(\frac{z_{41}^*}{\rho_1}\right)^3
-m_2^2+m_1^2
\end{array}
\end{equation}
\begin{equation}
\begin{array}{l}
\left(z_{12}-z_{11}^*\right)\left(z_{12}-z_{21}^*\right)
\left(z_{12}-z_{31}^*\right)\left(z_{12}-z_{41}^*\right)=
\\[3mm]\rule{20mm}{0mm}=
-i\Gamma_2m_2\left(\frac{z_{12}}{\rho_2}\right)^3
-i\Gamma_1m_1\left(\frac{z_{12}}{\rho_1}\right)^3
-m_1^2+m_2^2
\end{array}
\end{equation}
\begin{equation}
\begin{array}{l}
\left(z_{22}^*-z_{11}\right)\left(z_{22}^*-z_{21}\right)
\left(z_{22}^*-z_{31}\right)\left(z_{22}^*-z_{41}\right)=
\\[3mm]\rule{20mm}{0mm}=
i\Gamma_2m_2\left(\frac{z_{22}^*}{\rho_2}\right)^3
+i\Gamma_1m_1\left(\frac{z_{22}^*}{\rho_1}\right)^3
-m_1^2+m_2^2
\end{array}
\end{equation}
\begin{equation}
\begin{array}{l}
\left(z_{32}^*-z_{11}\right)\left(z_{32}^*-z_{21}\right)
\left(z_{32}^*-z_{31}\right)\left(z_{32}^*-z_{41}\right)=
\\[3mm]\rule{20mm}{0mm}=
i\Gamma_2m_2\left(\frac{z_{32}^*}{\rho_2}\right)^3
+i\Gamma_1m_1\left(\frac{z_{32}^*}{\rho_1}\right)^3
-m_1^2+m_2^2
\end{array}
\end{equation}
\begin{equation}
\begin{array}{l}
\left(z_{42}^*-z_{11}\right)\left(z_{42}^*-z_{21}\right)
\left(z_{42}^*-z_{31}\right)\left(z_{42}^*-z_{41}\right)=
\\[3mm]\rule{20mm}{0mm}=
i\Gamma_2m_2\left(\frac{z_{42}^*}{\rho_2}\right)^3
+i\Gamma_1m_1\left(\frac{z_{42}^*}{\rho_1}\right)^3
-m_2^2+m_1^2
\end{array}
\end{equation}

Here we can shorten the expressions using notations:
\begin{equation}
G=\frac{\Gamma_1m_1}{\rho_1^3}+\frac{\Gamma_2m_2}{\rho_2^3},\;\;\;
D_m^2=m_2^2-m_1^2.
\end{equation}

For the rest of free parameters $A_x=a_xe^{i\psi_x}$,
$B_x=b_xe^{-i\psi_x}$
there is a system of eight
complex equations:

\begin{equation}
\begin{array}{l}
a_xe^{i\psi_x}\left(\frac{a_xe^{-i\psi_x}}{
-2i\Gamma_1m_1\left(\frac{z_{11}}{\rho_1}\right)^3}+
\frac{b_xe^{i\psi_x}}{
-iGz_{11}^3-D_m^2}\right)=
\\[4mm]\rule{20mm}{0mm}
=ae^{i\psi}\left(\frac{ae^{-i\psi}}{
-2i\Gamma_1m_1\left(\frac{z_{11}}{\rho_1}\right)^3}+
\frac{be^{i\psi}}{
-iGz_{11}^3-D_m^2}\right),
\end{array}
\end{equation}

\begin{equation}
\begin{array}{l}
a_xe^{-i\psi_x}
\left(\frac{a_xe^{i\psi_x}}{
2i\Gamma_1m_1\left(\frac{z_{21}^*}{\rho_1}\right)^3
}+
\frac{b_xe^{-i\psi_x}}{
iGz_{21}^{*3}-D_m^2
}\right)
=\\[4mm]\rule{20mm}{0mm} =a e^{-i\psi}
\left(\frac{ae^{i\psi}}{
2i\Gamma_1m_1\left(\frac{z_{21}^*}{\rho_1}\right)^3
}+
\frac{be^{-i\psi}}{
iGz_{21}^{*3}-D_m^2
}\right)
\end{array}
\end{equation}

\begin{equation}
\begin{array}{l}
a_xe^{-i\psi_x}
\left(\frac{a_xe^{i\psi_x}}{
2i\Gamma_1m_1\left(\frac{z_{31}^*}{\rho_1}\right)^3
}+
\frac{b_xe^{-i\psi_x}}{
iGz_{31}^{*3}-D_m^2
}\right)=
\\[4mm]\rule{20mm}{0mm} =
ae^{-i\psi}
\left(\frac{ae^{i\psi}}{
2i\Gamma_1m_1\left(\frac{z_{31}^*}{\rho_1}\right)^3
}+
\frac{be^{-i\psi}}{
iGz_{31}^{*3}-D_m^2
}\right)
\end{array}
\end{equation}

\begin{equation}
\begin{array}{l}
a_xe^{-i\psi_x}
\left(\frac{a_xe^{i\psi_x}}{
2i\Gamma_1m_1\left(\frac{z_{41}^*}{\rho_1}\right)^3
}+
\frac{b_xe^{-i\psi_x}}{
iGz_{41}^{*3}-D_m^2
}\right)
= \\[4mm]\rule{20mm}{0mm} =
ae^{-i\psi}
\left(\frac{ae^{i\psi}}{
2i\Gamma_1m_1\left(\frac{z_{41}^*}{\rho_1}\right)^3
}+
\frac{be^{-i\psi}}{
iGz_{41}^{*3}-D_m^2
}
\right)
\end{array}
\end{equation}

\begin{equation}
\begin{array}{l}
b_xe^{-i\psi_x}\left(
\frac{a_xe^{-i\psi_x}}{-iGz_{12}^3+D_m^2}+
\frac{b_xe^{i\psi_x}}{
-2i\Gamma_2m_2\left(\frac{z_{12}}{\rho_2}\right)^3}
\right)
=\\[4mm]\rule{20mm}{0mm} =
be^{-i\psi}\left(
\frac{ae^{-i\psi}}{-iGz_{12}^3+D_m^2}+
\frac{be^{i\psi}}{
-2i\Gamma_2m_2\left(\frac{z_{12}}{\rho_2}\right)^3}
\right)
\end{array}
\end{equation}

\begin{equation}
\begin{array}{l}
b_xe^{i\psi_x}\left(
\frac{a_xe^{i\psi_x}}{iGz_{22}^{*3}+D_m^2}+
\frac{b_xe^{-i\psi_x}}{
2i\Gamma_2m_2\left(\frac{z_{22}^*}{\rho_2}\right)^3}
\right)
=\\[4mm]\rule{20mm}{0mm} =
be^{i\psi}\left(
\frac{ae^{i\psi}}{iGz_{22}^{*3}+D_m^2}+
\frac{be^{-i\psi}}{
2i\Gamma_2m_2\left(\frac{z_{22}^*}{\rho_2}\right)^3}
\right)
\end{array}
\end{equation}

\begin{equation}
\begin{array}{l}
b_xe^{i\psi_x}\left(
\frac{a_xe^{i\psi_x}}{iGz_{32}^{*3}+D_m^2}+
\frac{b_xe^{-i\psi_x}}{
2i\Gamma_2m_2\left(\frac{z_{32}^*}{\rho_2}\right)^3}
\right)
=\\[4mm]\rule{20mm}{0mm} =
be^{i\psi}\left(
\frac{ae^{i\psi}}{iGz_{32}^{*3}+D_m^2}+
\frac{be^{-i\psi}}{
2i\Gamma_2m_2\left(\frac{z_{32}^*}{\rho_2}\right)^3}
\right)
\end{array}
\end{equation}

\begin{equation}
\begin{array}{l}
b_xe^{i\psi_x}\left(
\frac{a_xe^{i\psi_x}}{iGz_{42}^{*3}+D_m^2}+
\frac{b_xe^{-i\psi_x}}{
2i\Gamma_2m_2\left(\frac{z_{42}^*}{\rho_2}\right)^3}
\right)
=\\[4mm]\rule{20mm}{0mm} =
be^{i\psi}\left(
\frac{ae^{i\psi}}{iGz_{42}^{*3}+D_m^2}+
\frac{be^{-i\psi}}{
2i\Gamma_2m_2\left(\frac{z_{42}^*}{\rho_2}\right)^3}
\right)
\end{array}
\end{equation}

Let us introduce new variables:
\begin{equation}
\xi=\frac{a_xb_x}{ab},\;\;\;
\nu_x=\frac{a_x}{b_x},\;\;\; \nu=\frac{a}{b}.
\end{equation}

Of course the trivial solution of the system of equations is valid:
 $\psi_x=\psi, \;
a_x=a,\; b_x=b,\; \xi=1,\; \nu_x=\nu$.

The new set of unknown variables $\psi_x,\; \nu_x,\; \xi$.
Old variables are connected with new ones with
the following relations:
$a_x=\sqrt{ab\xi\nu_x},\; b_x=\sqrt{\frac{ab\xi}{\nu_x}}$.

The system of equations with new unknowns  $\xi,\; \nu_x$ is:
\begin{equation}\label{eq:Final_system}
\left\{
\begin{array}{l}
e^{2i\psi_x}=\frac{e^{2i\psi}}{\xi}+
\frac{iGz_{11}^3+D_m^2}{2i\Gamma_1m_1\left(\frac{z_{11}}{\rho_1}\right)^3}
\cdot\left(\frac{\nu}{\xi}-\nu_x\right),\\
e^{-2i\psi_x}=\frac{e^{-2i\psi}}{\xi}+
\frac{iGz_{21}^{*3}-D_m^2}{2i\Gamma_1m_1\left(\frac{z_{21}^*}{\rho_1}\right)^3}
\cdot\left(\frac{\nu}{\xi}-\nu_x\right),\\
e^{-2i\psi_x}=\frac{e^{-2i\psi}}{\xi}+
\frac{iGz_{31}^{*3}-D_m^2}{2i\Gamma_1m_1\left(\frac{z_{31}^*}{\rho_1}\right)^3}
\cdot\left(\frac{\nu}{\xi}-\nu_x\right),\\
e^{-2i\psi_x}=\frac{e^{-2i\psi}}{\xi}+
\frac{iGz_{41}^{*3}-D_m^2}{2i\Gamma_1m_1\left(\frac{z_{41}^*}{\rho_1}\right)^3}
\cdot\left(\frac{\nu}{\xi}-\nu_x\right),\\
e^{-2i\psi_x}=\frac{e^{-2i\psi}}{\xi}+
\frac{iGz_{12}^{3}-D_m^2}{2i\Gamma_2m_2\left(\frac{z_{12}}{\rho_2}\right)^3}
\cdot\left(\frac{1}{\xi\nu}-\frac{1}{\nu_x}\right),\\
e^{2i\psi_x}=\frac{e^{2i\psi}}{\xi}+
\frac{iGz_{22}^{*3}+D_m^2}{2i\Gamma_2m_2\left(\frac{z_{22}^*}{\rho_2}\right)^3}
\cdot\left(\frac{1}{\xi\nu}-\frac{1}{\nu_x}\right),\\
e^{2i\psi_x}=\frac{e^{2i\psi}}{\xi}+
\frac{iGz_{32}^{*3}+D_m^2}{2i\Gamma_2m_2\left(\frac{z_{32}^*}{\rho_2}\right)^3}
\cdot\left(\frac{1}{\xi\nu}-\frac{1}{\nu_x}\right),\\
e^{2i\psi_x}=\frac{e^{2i\psi}}{\xi}+
\frac{iGz_{42}^{*3}+D_m^2}{2i\Gamma_2m_2\left(\frac{z_{42}^*}{\rho_2}\right)^3}
\cdot\left(\frac{1}{\xi\nu}-\frac{1}{\nu_x}\right).
\end{array}\right.
\end{equation}
Subtracting the third equation from the second one, we get
\begin{equation}\label{eq:nux=nu/xi}
\left(
\frac{iGz_{21}^{*3}-D_m^2}{2i\Gamma_1m_1\left(\frac{z_{21}^*}{\rho_1}\right)^3}
-
\frac{iGz_{31}^{*3}-D_m^2}{2i\Gamma_1m_1\left(\frac{z_{31}^*}{\rho_1}\right)^3}
\right)\cdot\left(\frac{\nu}{\xi}-\nu_x\right)=0,
\end{equation}
which can be only valid if one of the multipliers equals zero.
So one of the solution is
\begin{equation}
\nu_x=\frac{\nu}{\xi}.
\end{equation}
It can be checked easily that the first multiplier cannot
be equal to zero for any resonance parameters. If it were so then
\begin{equation}
iG-\frac{D_m^2}{z_{21}^{*3}}=iG-\frac{D_m^2}{z_{31}^{*3}},
\end{equation}
and this is equivalent to the equality $z_{21}=z_{31}$.
Similarly using the equation pairs ``second -- fourth'',
``third -- fourth'', and supposing $\nu_x\neq \frac{\nu}{\xi}$,
one gets $z_{21}=z_{31}=z_{41}$.
So the three roots of quatric equation are equal to the same value.
\begin{equation}
\begin{array}{l}
\rho^4+4\mu^2-m_1^2+i\Gamma_1m_1\left(\frac{\rho}{\rho_1}\right)^3=
\left(\rho-z_{11}\right)\left(\rho-z_{21}\right)^3=
\\[3mm] =
\rho^4-\left(z_{11}+3z_{21}\right)\rho^3+
\\[3mm] \rule{20mm}{0mm}+
3z_{21}\left(z_{11}+z_{21}\right)\rho^2-
z_{21}^2\left(3z_{11}+z_{21}\right)\rho+z_{11}z_{21}^3.
\end{array}
\end{equation}
So as none of the roots is equal to zero, then
comparing to representations of the same equation one concludes that
\begin{equation}
\left\{
\begin{array}{l}
z_{11}+z_{21}=0,\\
3z_{11}+z_{21}=0,
\end{array} \right.
\end{equation}
and hence $z_{11}=z_{21}=0$, which is impossible.
So our guess $\nu_x\neq \frac{\nu}{\xi}$
is invalid and the solution of~(\ref{eq:nux=nu/xi}) is
\begin{equation}
\nu_x=\frac{\nu}{\xi}.
\end{equation}

Similar analysis of equations 6--8
in (\ref{eq:Final_system})
gives
\begin{equation}
\nu_x=\xi\nu.
\end{equation}
So one can immediately conclude that
\begin{equation}
\xi=1,\;\; \nu_x=\nu.
\end{equation}
Now from any equation from the system~(\ref{eq:Final_system})
one derives
\begin{equation}
e^{2i\psi_x}=e^{2i\psi}\Longrightarrow \psi_x=\psi.
\end{equation}

As a result we conclude that for the case of energy dependent
resonance width the degenaration disappears and there is
the only set of resonance parameters presenting the given energy
dependence of cross section.

\section{Numeric experiments}
For additional check of these conclusions it is useful to
carry out numeric experiments simulating some actual experiment in high energy
physics.
Such experiments can show the role of experimental statistics.

In all cases we should suppose some true process cross section function
 $\sigma(E)$, where  $E$ is the energy of colliding beams,
then for the finite number of points $E_k$ we generate
experimental number of events, according to the Poisson probability distribution
(integrated luminosity at every point is equal to the same value $L$).

In order to get the parameters of resonances, consisting the given model
of process, we minimize the likelihood function as follows
\begin{equation}
{\cal L}=\sum\limits_{k=1}^N2\left(p_k-n_k+n_k\ln\frac{n_k}{p_k}\right),
\end{equation}
which is equal to the doubled logarithmic likelihood function with
opposite sign.
Here  $p_k=\sigma(E_k)\cdot L$.
For greater statistics this function limits to
 $\chi^2$ value, that can be used for the check of statistical
confidence level.
Minimization will be performed with well-known MINUIT package~\cite{MINUIT}.

For presentation of experimental cross section on the plots,
the experimental number of events will be ascribed asymmetric statistical errors:
\begin{equation}
\Delta n_i^{(+)}=\sqrt{n_i+1},\;\; \Delta n_i^{(-)}=\sqrt{n_i}.
\end{equation}

\subsection{Approximate expression for the resonance amplitude
}

\begin{equation}\label{eq:FirstExperiment}
\sigma(E)=\left|\frac{a}{E-m_1+i\frac{\Gamma_1}{2}}+
\frac{be^{i\psi}}{E-m_2+i\frac{\Gamma_2}{2}}\right|^2
\end{equation}
with ``true'' values of parameters
\begin{equation}
\begin{array}{l}
m_1=782.6,\;\;\Gamma_1=8.4,\;\;\; a=1,\\
m_2=1019.4,\;\;\Gamma_2=4.5,\;\; b=0.3,\;\;
\psi=155^\circ.
\end{array}
\end{equation}

\begin{figure}[tbp]
\epsfxsize=0.48\textwidth
\epsfbox{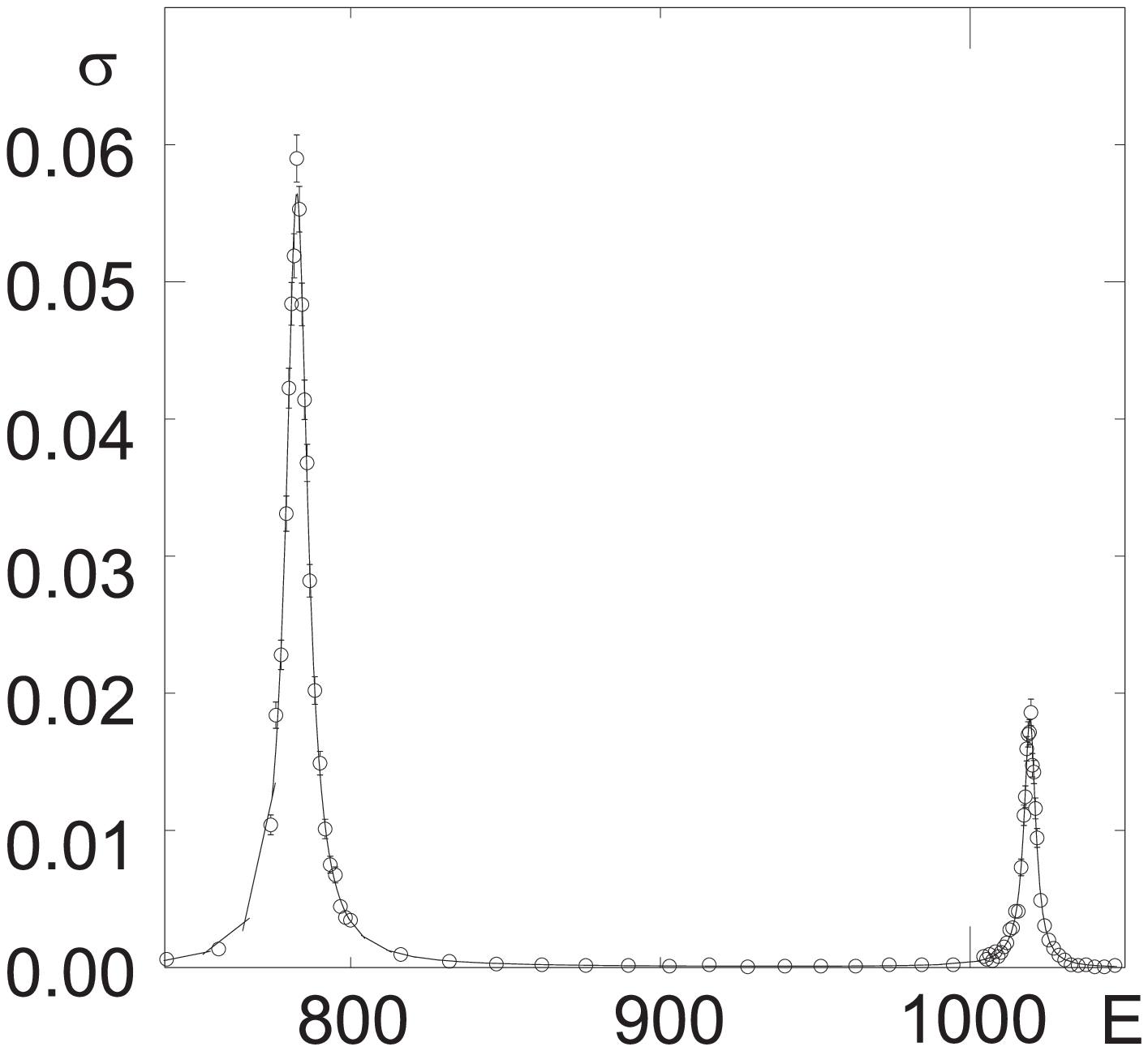}
\hfill
\epsfxsize=0.48\textwidth
\epsfbox{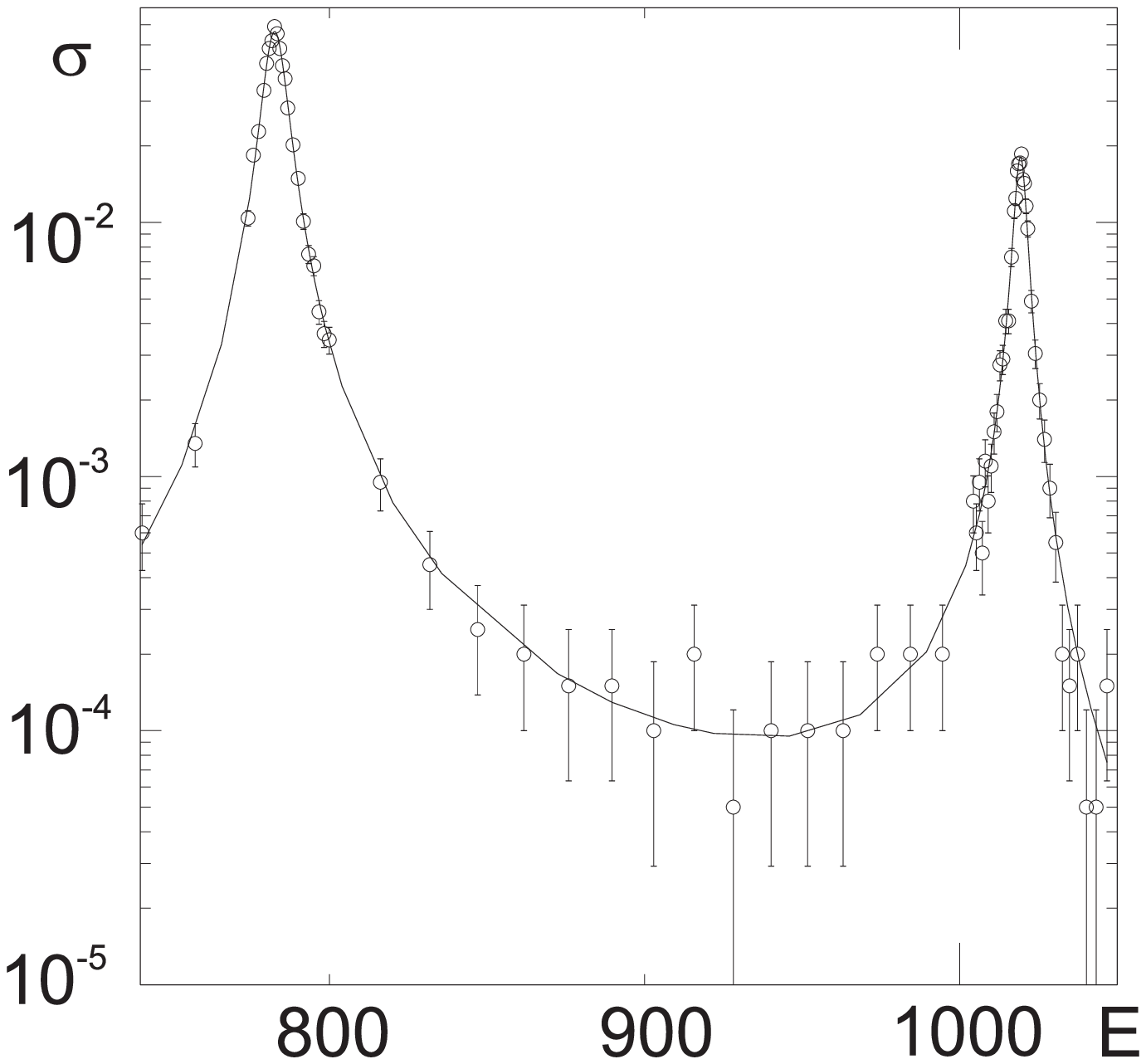}
\caption{\label{FitExp1}Result of fit to ``experimental'' points.
Cross section model is~(\ref{eq:FirstExperiment}).
$\chi^2/n_D=64.3/(74-7)$.}
\end{figure}

This function is equal to $\sigma(m_1)\approx \frac{1}{4.2^2}=0.057$,
$\sigma(m_2)\approx \frac{0.09}{2.25^2}=0.018$.
In order to an accuracy at the resonance maxima to be at least at the 5\% level let us appoint $L=2\cdot  10^4$.
Fig.\,\ref{FitExp1}
shows the ``experimental'' set of points.

\begin{figure}[tbp]
\epsfxsize=0.45\textwidth
\centerline{\epsfbox{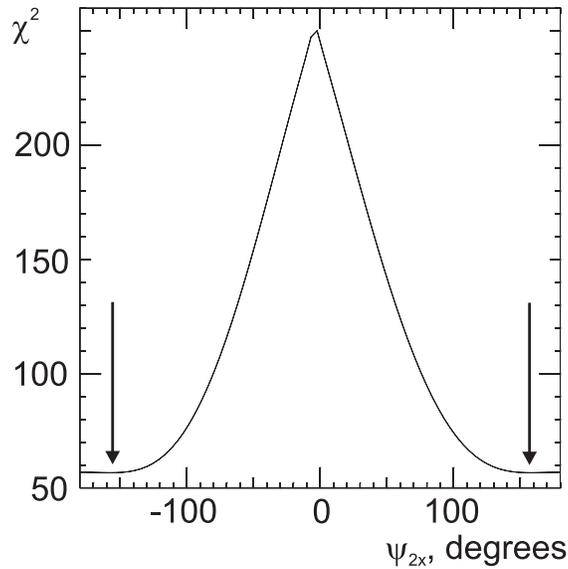}}
\caption{\label{Fit1_results}Plot of the likelihood function
on the
phase of the second resonance amplitude $\psi_{2x}$
}
\end{figure}

Plot of likelihood function 
on
the phase of second resonance amplitude $\psi_{2x}$  is
presented in Fig\,\ref{Fit1_results}.
Two equivalent minima with $\chi^2/n_D=56.703/(74-7)$
are obtained at $\psi_{2x}=-157.14^\circ$ and $\psi_{2x}=157.51^\circ$.
Although the separating maximum is not high ($\chi^2=56.907$ at
$\psi_{2x}=\pm 180^\circ$), these are the different solutions.
Values of all parameters at these minimum points are cited at the Table\,\ref{Fit1_2min}.
\begin{table}[tbp]
\caption{\label{Fit1_2min}Parameters of resonances at the minimum points
of likelihood function.}
\begin{center}
\begin{tabular}{|c|c|c|c|c|c|c|c|}
\hline
$a_x$ & $m_{1x}$ & $\Gamma_{1x}$ & $b_{x}$ & $m_{2x}$ & $\Gamma_{2x}$ &
$\psi_{2x}^\circ$ & $\chi^2$ \\
\hline
1.0125 & 782.62 & 8.5371 & 0.30284 & 1019.4 & 4.5554 & 157.815 & 56.703 \\
\hline
1.0167 & 782.62 & 8.5371 & 0.31033 & 1019.4 & 4.5554 & -157.065 & 56.703 \\
\hline
\end{tabular}
\end{center}
\end{table}

This numeric experiment confirmed the analytical conclusion ---
fitting the data with the approximate resonance formula produce the ambiguity
of resonance phases and amplitudes.

\subsection{Relativistic form of resonance amplitude}

Let us consider more accurate dependence of the resonance amplitude
on energy:

\begin{equation}\label{eq:SecondExperiment}
\sigma(E)=\frac{m_1^4\sqrt{E^2-4\mu^2}^3}{E^4\sqrt{m_1^2-4\mu^2}^3}\cdot
\left|\frac{2m_1a}{E^2-m_1^2+i{\Gamma_1}{m_1}}+
\frac{2m_2be^{i\psi}}{E^2-m_2^2+i{\Gamma_2}{m_2}}\right|^2
\end{equation}
with the same ``true'' parameters values
\begin{equation}
\begin{array}{l}
m_1=782.6,\;\;\Gamma_1=8.4,\;\;\; a=1,\\
m_2=1019.4,\;\;\Gamma_2=4.5,\;\; b=0.3,\;\;
\psi=155^\circ,
\end{array}
\end{equation}
where the factors $2m_i$ are introduced in order to keep
the valeus of amplitudes at the resonance masses, and general
factor imitates the threshold behaviour of
cross section with  $\mu\approx \frac{3m_\pi}{2}\approx
\frac{140+140+135}{2}\approx 208$.
In order that minima of likelihood function to be more demonstrative
let us increase the integrated luminosity to the value of $L=10^6$.

\begin{figure}[tbp]
\epsfxsize=0.48\textwidth
\epsfbox{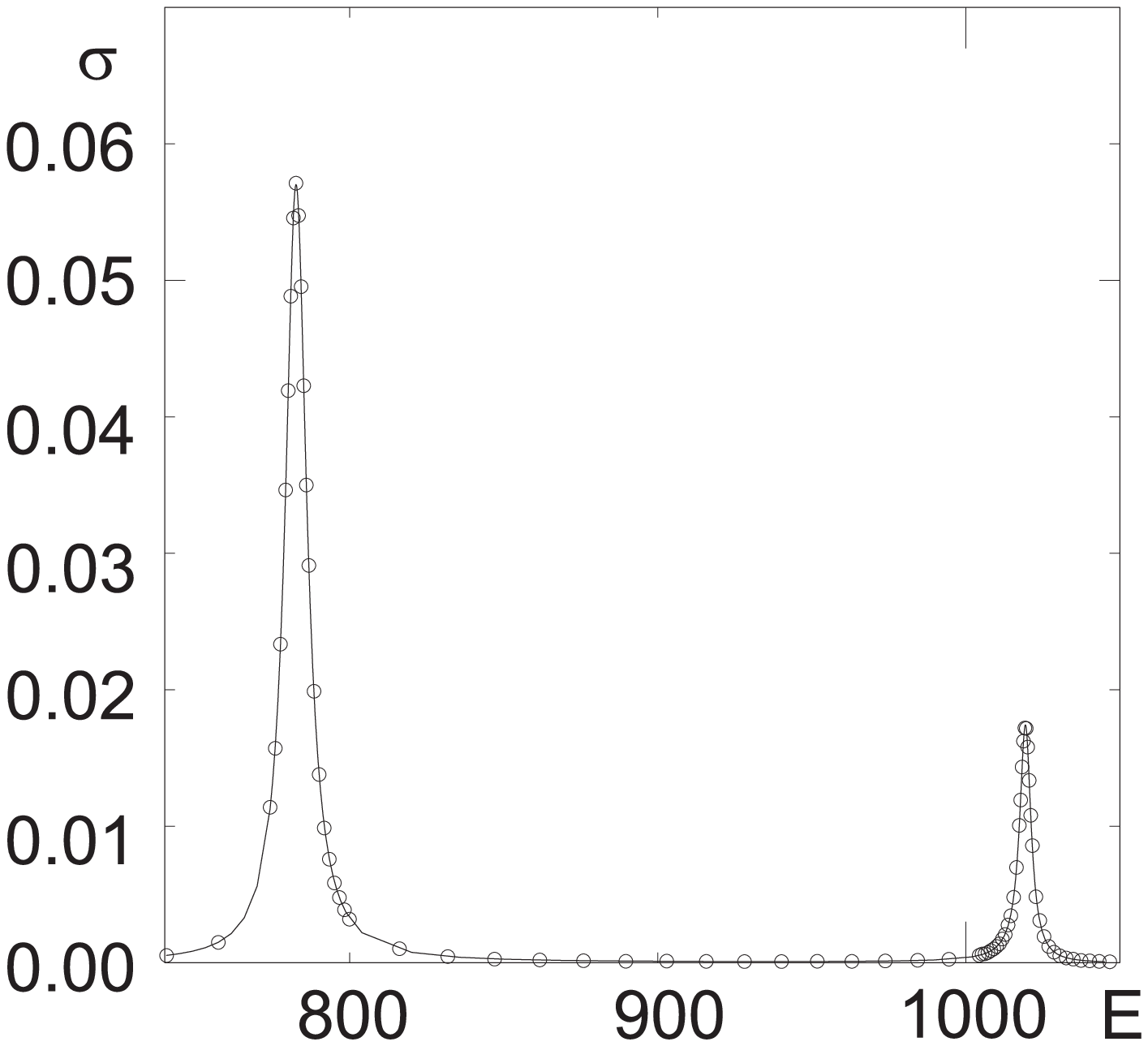}
\hfill
\epsfxsize=0.48\textwidth
\epsfbox{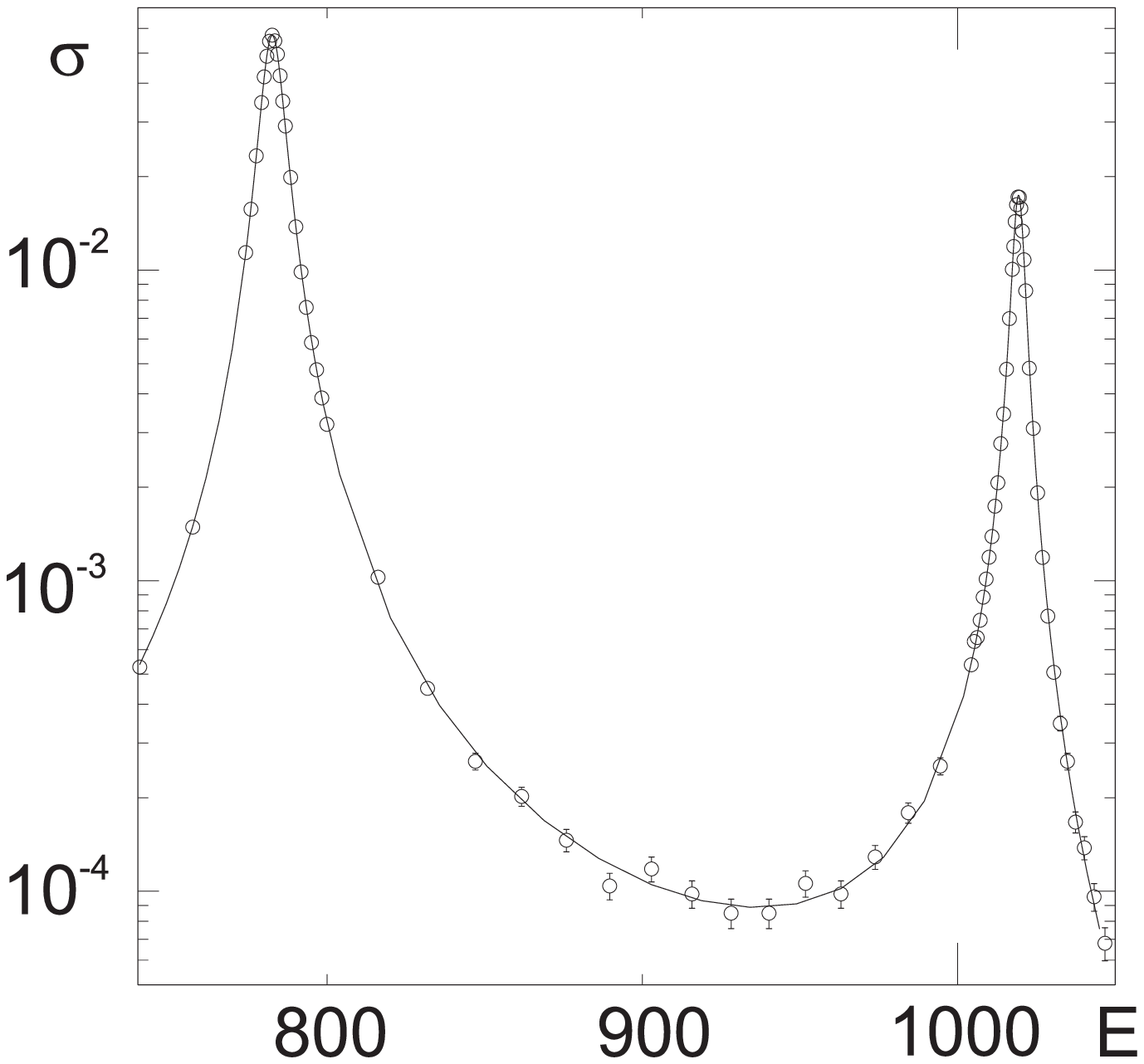}
\caption{\label{FitExp2}Result of fit of ``experimental'' points.
Cross section model is described with the formula~(\ref{eq:SecondExperiment}).
$\chi^2/n_D=74.6/(74-7)$.}
\end{figure}

In Fig.\,\ref{FitExp2}
the set of points and optimal cross section of the type~(\ref{eq:SecondExperiment})
are presented.

\begin{figure}[tbp]
\epsfxsize=0.45\textwidth
\centerline{\epsfbox{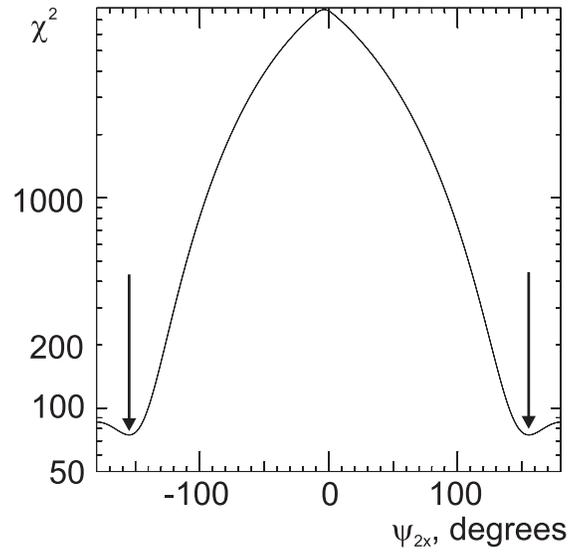}}
\caption{\label{Fit2_results}Plot of likelihood function
on the second resonance phase $\psi_{2x}$
}
\end{figure}

The plot of the likelihood function
on the second resonance phase  $\psi_{2x}$
is presented in Fig.\,\ref{Fit2_results}.

As we expected, we get two equivalent minima again.
Parameters values at the minimum points of $\cal L$
are shown in Table\,\ref{Fit2_2min}.
\begin{table}[tbp]
\caption{\label{Fit2_2min}Resonances parameters at the two minimum points of likelihood
function.}
\begin{center}
\begin{tabular}{|c|c|c|c|c|c|c|c|}
\hline
$a_x$ & $m_{1x}$ & $\Gamma_{1x}$ & $b_{x}$ & $m_{2x}$ & $\Gamma_{2x}$ &
$\psi_{2x}^\circ$ & $\chi^2$ \\
\hline
1.0020 & 782.60 & 8.4116 & 0.30007 & 1019.4 & 4.5093 & 155.283 & 74.58624 \\
\hline
1.0071 & 782.60 & 8.4116 & 0.30708 & 1019.4 & 4.5093 & -154.807 & 74.58624 \\
\hline
\end{tabular}
\end{center}
\end{table}

\subsection{Resonance width dependent on energy}
Finally let us consider the case where the degeneracy is expected to disappear
and only one global minimum will be found:

\begin{equation}\label{eq:ThirdExperiment}
\begin{array}{l}
\sigma(E)=\frac{m_1^4\sqrt{E^2-4\mu^2}^3}{E^4\sqrt{m_1^2-4\mu^2}^3}\times
\\[4mm] \rule{3mm}{0mm}\times
\left|\frac{2m_1a}{E^2-m_1^2+i{\Gamma_1}{m_1}
\cdot\left(\frac{E^2-4\mu^2}{m_1^2-4\mu^2}
\right)^{\frac{3}{4}}}+
\frac{2m_2be^{i\psi}}{E^2-m_2^2+i{\Gamma_2}{m_2}
\cdot\left(\frac{E^2-4\mu^2}{m_2^2-4\mu^2}
\right)^{\frac{3}{4}}}\right|^2
\end{array}
\end{equation}

\begin{figure}[tbp]
\epsfxsize=0.48\textwidth
\epsfbox{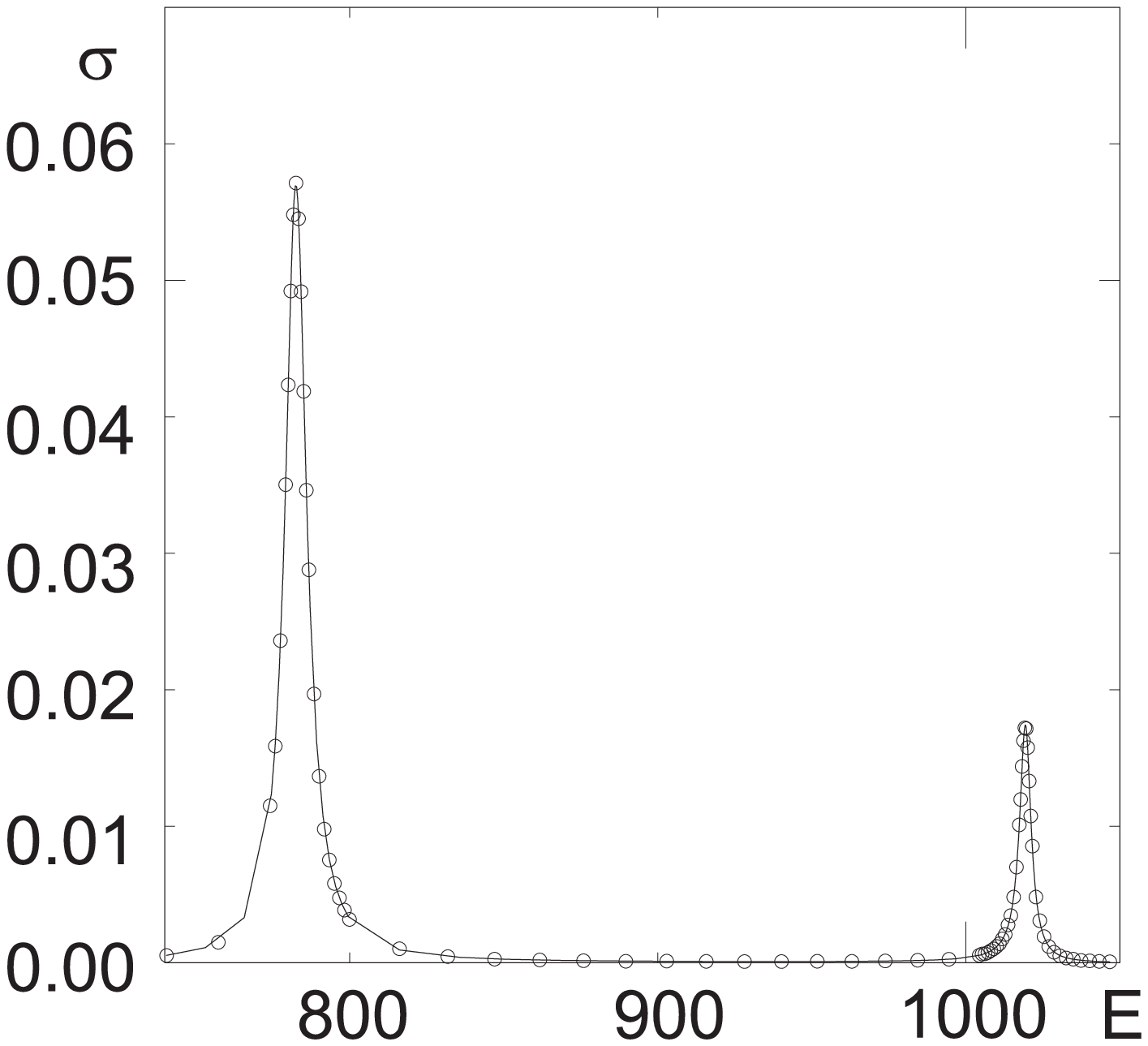}
\hfill
\epsfxsize=0.48\textwidth
\epsfbox{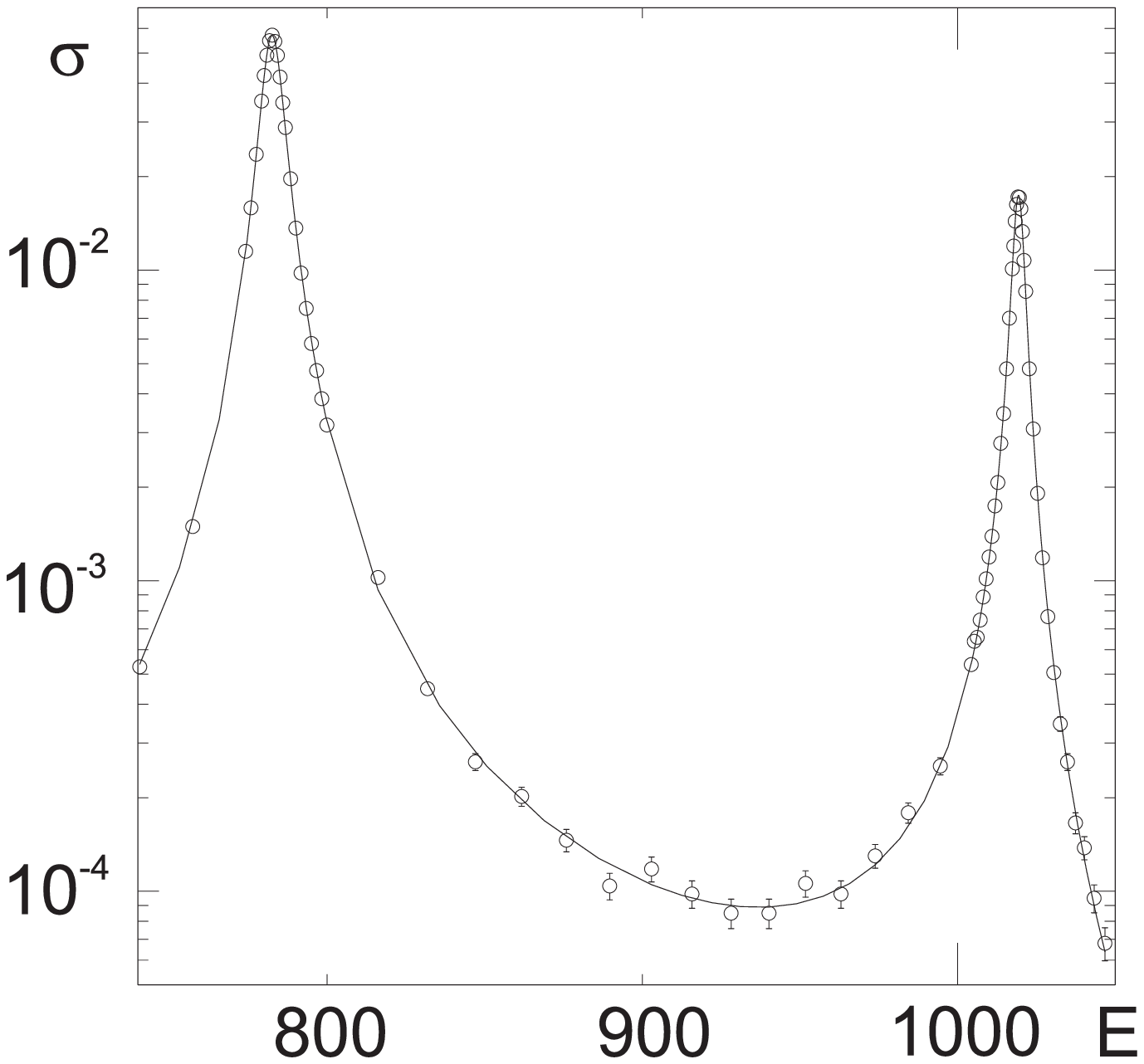}
\caption{\label{FitExp3}Result of fit of ``experimental'' points.
Cross section is described by the formula~(\ref{eq:ThirdExperiment}).
$\chi^2/n_D=74.8/(74-7)$.}
\end{figure}

Fig.\,\ref{FitExp3}
shows the set of energy points and optimal cross section of the type~(\ref{eq:ThirdExperiment}).

\begin{figure}[tbp]
\epsfxsize=0.45\textwidth
\centerline{\epsfbox{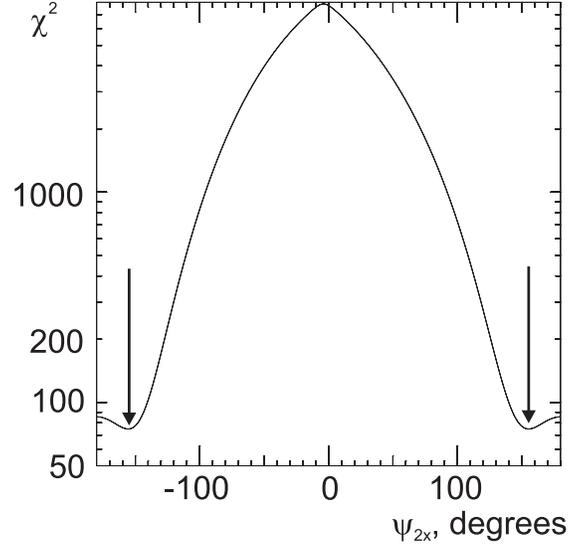}}
\caption{\label{Fit3_results}Likelihood function plot
on the phase of the second resonance  $\psi_{2x}$
 }
\end{figure}

Likelihood function plot
on the phase of the second resonance  $\psi_{2x}$
is shown in Fig.\,\ref{Fit3_results}.

Indeed, the minimum values are not equal now (see Table\,\ref{Fit3_2min}),
however the difference is very small and, what was totally
unexpected, the ``better minimum'' corresponds to the ``wrong'' minimum.
\begin{table}[tbp]
\caption{\label{Fit3_2min}Resonance parameters at the two points
of minimum of likelihood function.}
\begin{center}
\begin{tabular}{|c|c|c|c|c|c|c|c|}
\hline
$a_x$ & $m_{1x}$ & $\Gamma_{1x}$ & $b_{x}$ & $m_{2x}$ & $\Gamma_{2x}$ &
$\psi_{2x}^\circ$ & $\chi^2$ \\
\hline
1.0020 & 782.60 & 8.4117 & 0.30007 & 1019.4 & 4.5093 & 155.289 & 74.82254 \\
\hline
1.0071 & 782.60 & 8.4117 & 0.30699 & 1019.4 & 4.5093 & -155.529 & 74.81663 \\
\hline
\end{tabular}
\end{center}
\end{table}

Let us look what will change if the experimental statistics will increase
by factor of 100
($L=10^8$).
In Table \,\ref{Fit4_2min} the parameters of resonances are shown
for two minimum points of likelihood function. Again
the difference at the minimum points is very small and again
``wrong'' minimum is a little preferrable although
the total statistics is extreamly high and practically unreachable
in real experiments.

\begin{table}[tbp]
\caption{\label{Fit4_2min}Resonance parameters at the two minima
of likelihood function.}
\begin{center}
\begin{tabular}{|c|c|c|c|c|c|c|c|}
\hline
$a_x$ & $m_{1x}$ & $\Gamma_{1x}$ & $b_{x}$ & $m_{2x}$ & $\Gamma_{2x}$ &
$\psi_{2x}^\circ$ & $\chi^2$ \\
\hline
0.99982 & 782.60 & 8.3977 & 0.30015 & 1019.4 & 4.5025 & 154.985 & 70.085 \\
\hline
1.0050 & 782.60 & 8.3977 & 0.30711 & 1019.4 & 4.5025 & -155.237 & 70.060 \\
\hline
\end{tabular}
\end{center}
\end{table}

Let us return to the previous level of statistics ($L=10^6$),
but change the threshold factor $\mu=350$ instead of  208.
Again two minima (Table\,\ref{Fit5_2min}) difference
is statistically unreliable.
\begin{table}[tbp]
\caption{\label{Fit5_2min}Resonance parameters at the minimum points
of likelihood function.}
\begin{center}
\begin{tabular}{|c|c|c|c|c|c|c|c|}
\hline
$a_x$ & $m_{1x}$ & $\Gamma_{1x}$ & $b_{x}$ & $m_{2x}$ & $\Gamma_{2x}$ &
$\psi_{2x}^\circ$ & $\chi^2$ \\
\hline
0.99946 & 782.61 & 8.3931 & 0.30013 & 1019.4 & 4.5202 & 151.335 & 63.799 \\
\hline
1.0051 & 782.61 & 8.3935 & 0.30777 & 1019.4 & 4.5203 & -153.908 & 63.711 \\
\hline
\end{tabular}
\end{center}
\end{table}

\begin{figure}[tbp]
\epsfxsize=0.48\textwidth
\epsfbox{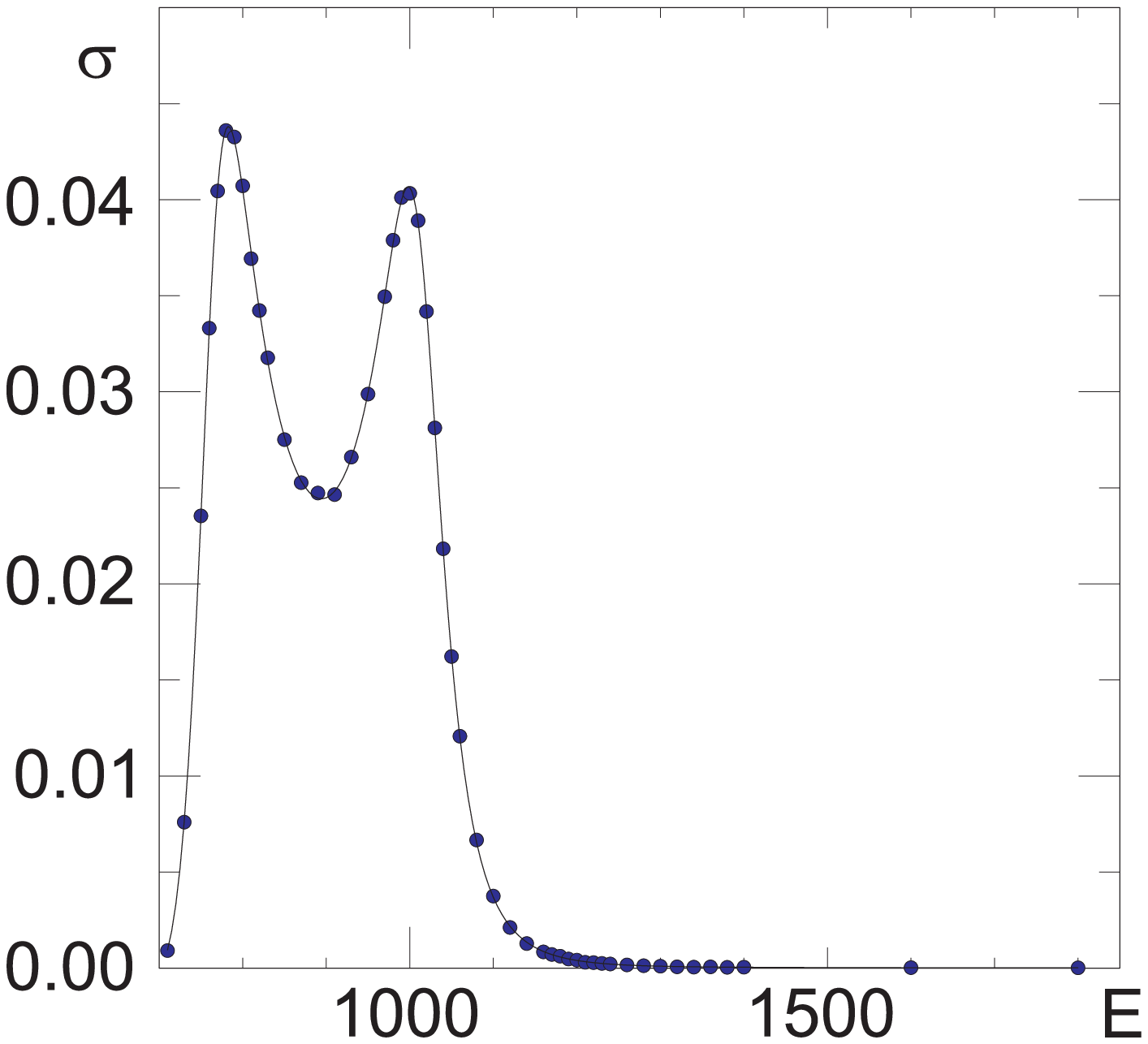}
\hfill
\epsfxsize=0.48\textwidth
\epsfbox{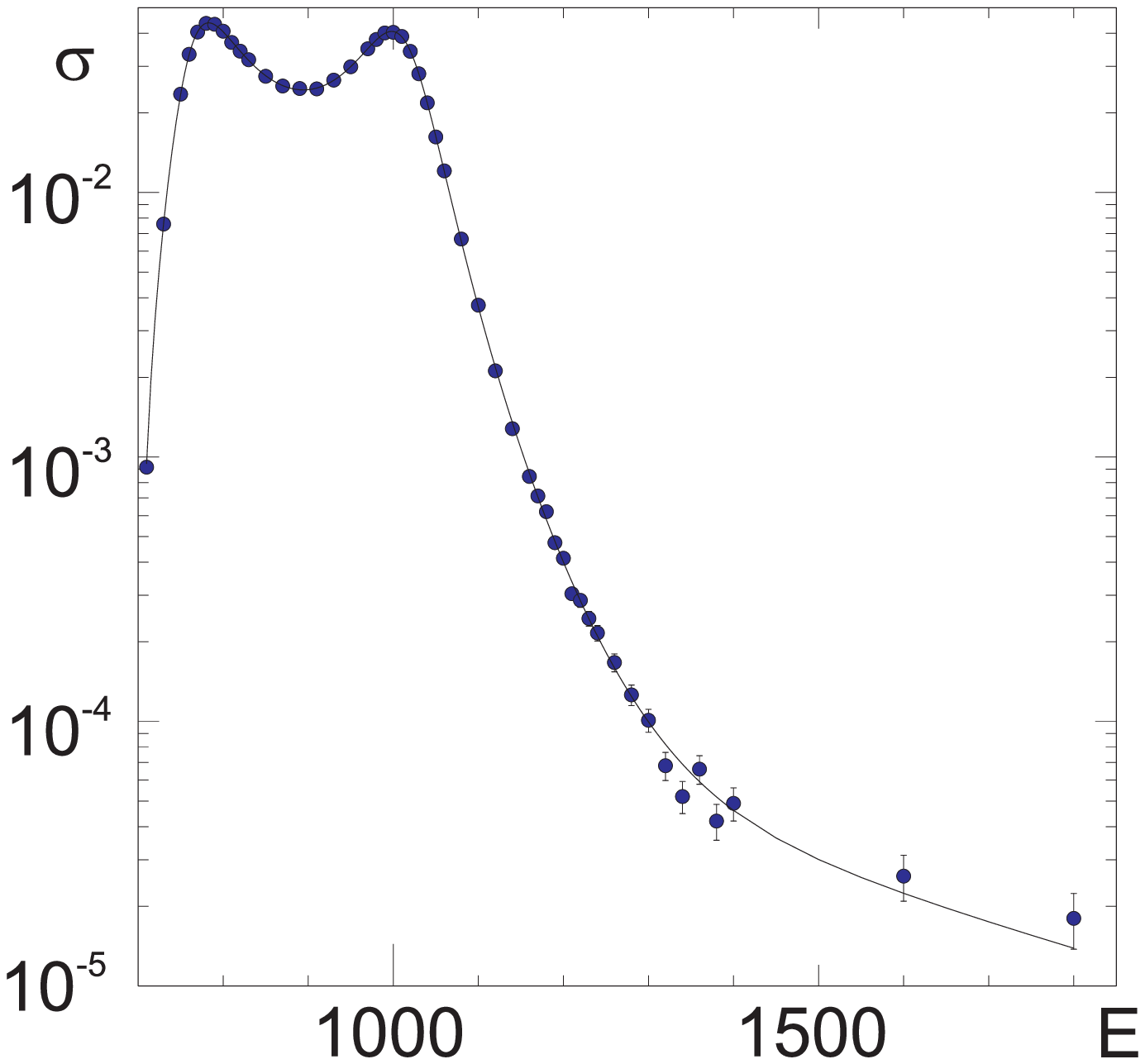}
\caption{\label{FitExp6}Fit result of ``data points''.
Cross section model is described by the formula~(\ref{eq:ThirdExperiment}).
$\chi^2/n_D=46.8/(50-7)$.}
\end{figure}

Evidently the more narrow are the resonances, the less is the influence
of width dependence on energy to the form of cross section.
Let us set both widthes large --- $\Gamma_{1,2}=100$.
In order that cross section at the resonance maxima to decrease not so much
let us take larger amplitudes: $a=10$, $b=5$.
Fig.~\ref{FitExp6} shows the ``data points'' and fit result.

\begin{figure}[tbp]
\epsfxsize=0.45\textwidth
\centerline{\epsfbox{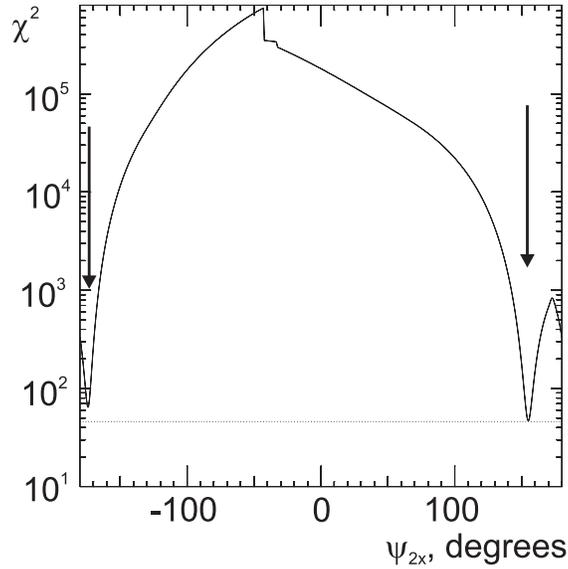}}
\caption{\label{Fit6_results}Likelihood function plot
on the second resonance
phase $\psi_{2x}$
}
\end{figure}

Likelihood function plot
on the second resonance phase $\psi_{2x}$ is presented in Fig.\,\ref{Fit6_results}.
This time the minimum values are essentially unequal and
the ``better'' minimum has ``correct'' phase.
Table\,\ref{Fit6_2min} shows the parameters of resonances
at these minimum points.
$\chi^2$ confidence level of the first minimum is $P_{50-7}(46.844)=0.318$,
the second one ---
$P_{50-7}(64.428)=0.0188$.

\begin{table}[tbp]
\caption{\label{Fit6_2min}Resonance parameters at the minimum points
of likelihood function.}
\begin{center}
\begin{tabular}{|c|c|c|c|c|c|c|c|}
\hline
$a_x$ & $m_{1x}$ & $\Gamma_{1x}$ & $b_{x}$ & $m_{2x}$ & $\Gamma_{2x}$ &
$\psi_{2x}^\circ$ & $\chi^2$ \\
\hline
10.050 & 782.71 & 100.61 & 5.0111 & 1019.4 & 100.22 & 154.985 & 46.844 \\
\hline
10.848 & 783.39 & 102.03 & 5.9767 & 1019.1 & 99.816 & -173.990 & 64.428 \\
\hline
\end{tabular}
\end{center}
\end{table}

\begin{figure}[tbp]
\epsfxsize=0.48\textwidth
\epsfbox{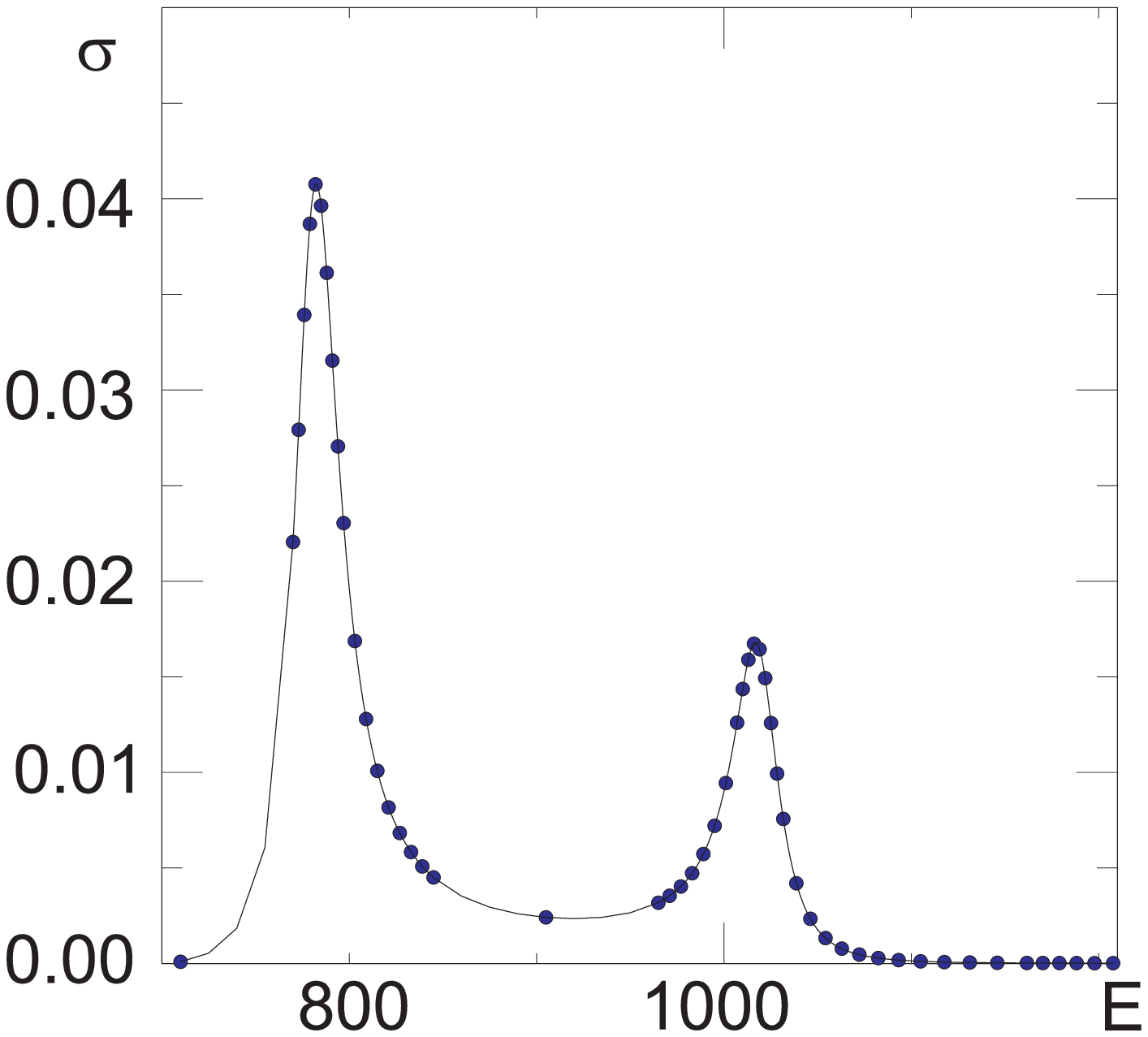}
\hfill
\epsfxsize=0.48\textwidth
\epsfbox{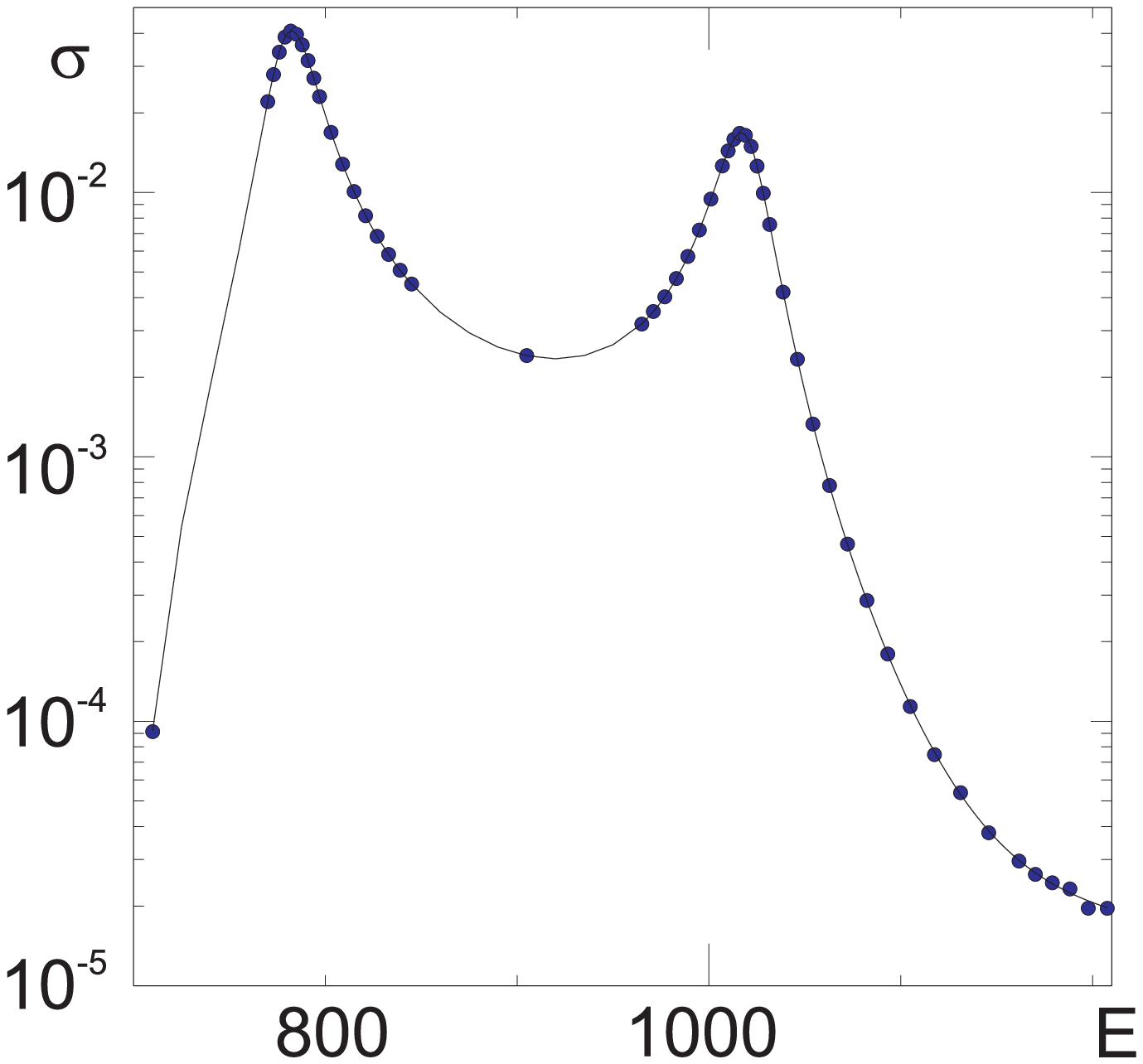}
\caption{\label{FitExp7}Result of fit of ``data points''.
Model of the cross section is described by the formula~(\ref{eq:ThirdExperiment}).
$\chi^2/n_D=52.05/(53-7)$.}
\end{figure}

For completeness let us consider the intermediate case:
$\Gamma_{1,2}=30$, $a=3$, $b=1$.
Fig.~\ref{FitExp7} demonstrates the ``data points''
and fit result.

Again two minima  (Table\,\ref{Fit7_2min})
are statistically equivalent (the difference of $\chi^2$ values much less
than unit).
\begin{table}[tbp]
\caption{\label{Fit7_2min}Resonance parameters at the two likelihood
minimum points.}
\begin{center}
\begin{tabular}{|c|c|c|c|c|c|c|c|}
\hline
$a_x$ & $m_{1x}$ & $\Gamma_{1x}$ & $b_{x}$ & $m_{2x}$ & $\Gamma_{2x}$ &
$\psi_{2x}^\circ$ & $\chi^2$ \\
\hline
3.0103 & 782.62 & 30.128 & 0.9986 & 1019.5 & 30.013 & 155.128 & 52.054 \\
\hline
3.0745 & 782.63 & 30.172 & 1.1311 & 1019.4 & 30.019 & -158.262 & 51.714 \\
\hline
\end{tabular}
\end{center}
\end{table}

After this numerical experiments we can conclude: if resonance width depends on energy,
then the two minimum points of likelihood function
describe not the same cross section function of energy.
However the difference can be used for cutting off the false minimum only
under the favourable conditions: high statistics, coverage of wide energy
interval with both resonances within, and the widths of the resonances
must be compatible with the mass difference.
The final result can be obtained only after comparison
the likelihood function values  $\chi^2$ at minima:
if the difference of levels is much greater than unit, then
one can choose better set of phases and amplitudes,
otherwise one should involve additional
considerations for the choice of interference phase.

\subsection{Three resonances}
\begin{figure}[tbp]
\epsfxsize=0.48\textwidth
\epsfbox{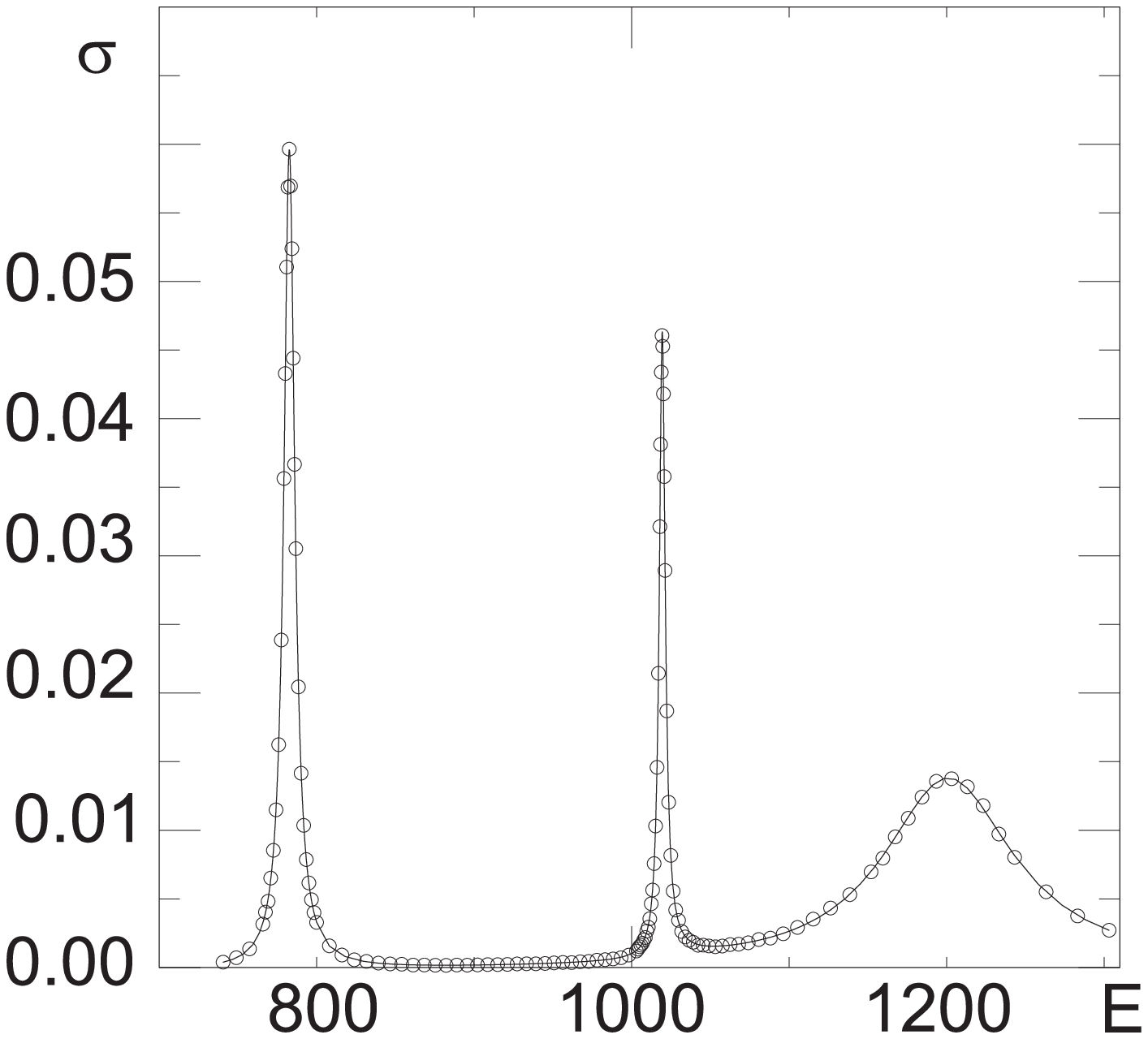}
\hfill
\epsfxsize=0.48\textwidth
\epsfbox{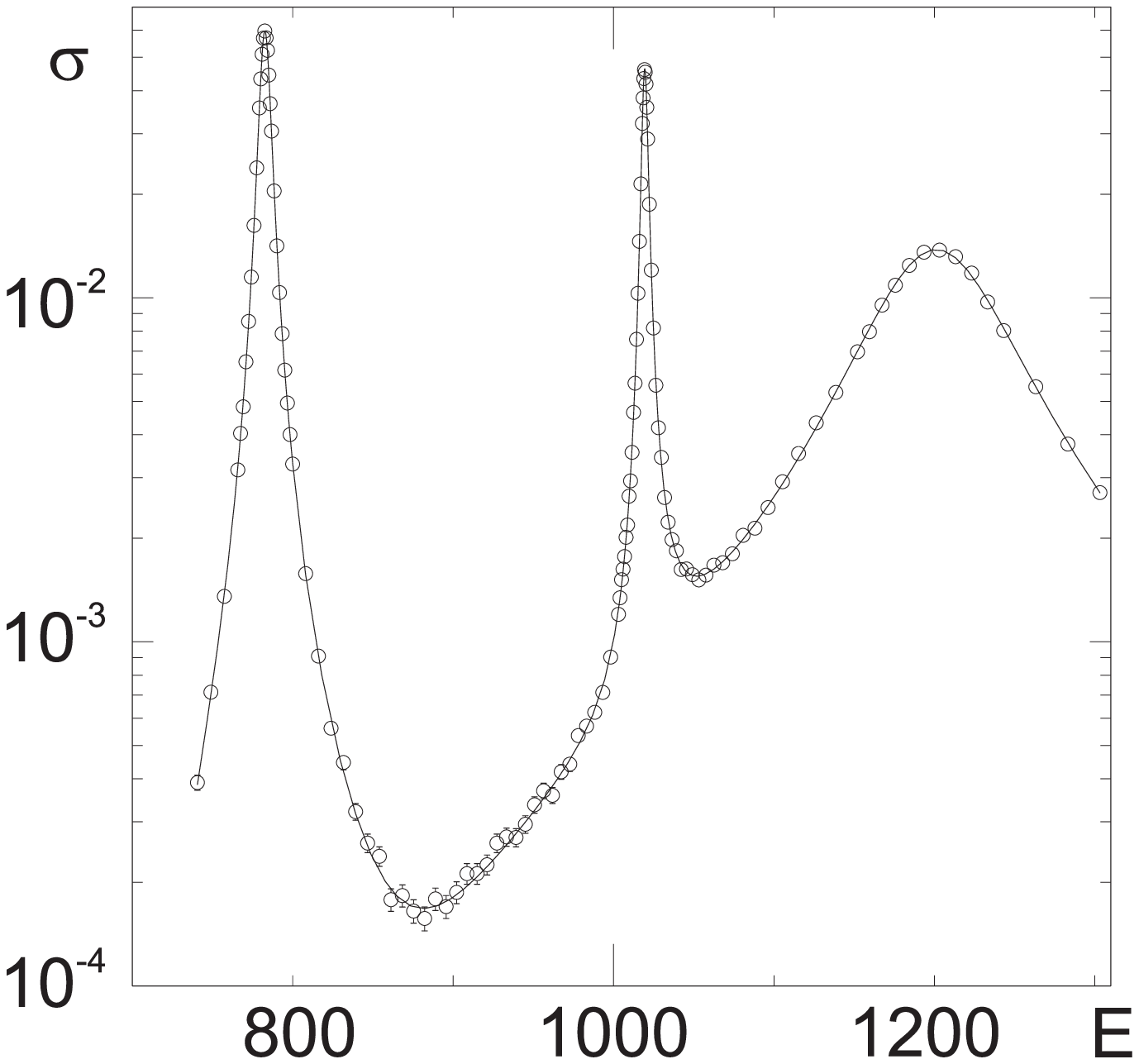}
\caption{\label{FitExp8}Result of the fit to ``data'' points.
Process cross section model is described by the formula~(\ref{eq:EighthExperiment}).
$\chi^2/n_D=126.0/(123-7)$.}
\end{figure}

The case of three resonances with constant widths:

\begin{equation}\label{eq:EighthExperiment}
\begin{array}{l}
\sigma(E)=\frac{m_1^4\sqrt{E^2-4\mu^2}^3}{E^4\sqrt{m_1^2-4\mu^2}^3}\cdot
\left|\frac{2m_1a}{E^2-m_1^2+i{\Gamma_1}{m_1}}+
\right. \\[4mm]\rule{40mm}{0mm}\left.
+\frac{2m_2be^{i\psi_2}}{E^2-m_2^2+i{\Gamma_2}{m_2}}
+\frac{2m_3ce^{i\psi_3}}{E^2-m_3^2+i{\Gamma_3}{m_3}}\right|^2
\end{array}
\end{equation}
with ``true'' values of parameters
\begin{equation}
\begin{array}{l}
m_1=782.6,\;\;\Gamma_1=8.4,\;\;\; a=1,\\
m_2=1019.4,\;\;\Gamma_2=4.5,\;\; b=0.3,\;\;
\psi_2=155^\circ,\\
m_3=1200,\;\;\Gamma_3=100,\;\; c=3,\;\;
\psi_3=30^\circ,
\end{array}
\end{equation}
$\mu=350$,
and integrated luminosity equals $L=10^6$.
Fig.~\ref{FitExp8} demonstrates ``experimental'' data and fit result.

\begin{figure}[tbp]
\epsfxsize=0.47\textwidth
\centerline{\epsfbox{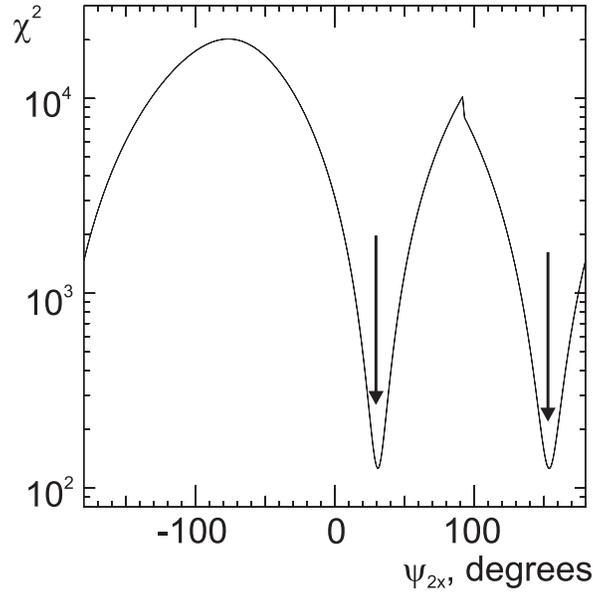}}
\caption{\label{Fit8_results}Likelihood function plot
on the phase of the second resonance phase $\psi_{2x}$
}
\end{figure}

Likelihood function plot vs the second
resonance phase $\psi_{2x}$ is shown in Fig.\,\ref{Fit8_results}.

One can see two minimum points on this plot (Table\,\ref{Fit8_2min}).

\begin{figure}[tbp]
\epsfxsize=0.47\textwidth
\centerline{\epsfbox{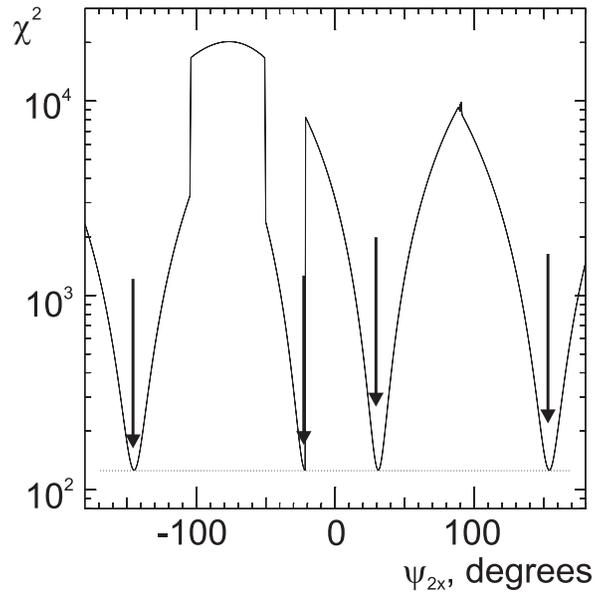}}
\caption{\label{Fit9_results}Likelihood function plot
on the phase of the second resonance phase $\psi_{2x}$
}
\end{figure}

\begin{table}[tbp]
\caption{\label{Fit8_2min}Resonances parameters at the two likelihood
function minimum  points.}
\begin{center}\footnotesize$\rule{0mm}{1mm}\!\!\!\!\!\!\!$%
\begin{tabular}
{|c|c|c|c|c|c|c|c|c|c|c|c|}
\hline
$a_x$ & $m_{1x}$ & $\Gamma_{1x}$
& $b_{x}$ & $m_{2x}$ & $\Gamma_{2x}$ & $\psi_{2x}^\circ$
& $c_{x}$ & $m_{3x}$ & $\Gamma_{3x}$ & $\psi_{3x}^\circ$
& $\chi^2$ \\
\hline
0.99805 & 782.62 & 8.3768 & 0.30038 & 1019.4 & 4.5055 & 153.925 &
3.0128 & 1199.7 & 100.75 & 28.634 & 125.969 \\
\hline
1.0535 & 782.62 & 8.3768 & 0.30323 & 1019.4 & 4.5055 & \mbox{30.931} &
 3.1673 & 1199.7 & 100.75 & \mbox{-72.318} &
 125.969 \\
\hline
\end{tabular}
\end{center}
\end{table}
Despite that we could not derive explicit analytical solutions
for the case of three resonances it seems that there are at least
two equivalent solutions. Let us check whether there are some more
solutions scanning the space of two parameters: $\psi_{2x}$
and $\psi_{3x}$.
All local minima are presented in Table\,\ref{List_ofScanned_min1}.
\begin{table}[tbp]
\caption{\label{List_ofScanned_min1}Resonance parameters at the likelihood
function local minimum points.}
\begin{center}\footnotesize$\rule{0mm}{1mm}\!\!\!\!\!\!\!$%
\begin{tabular}
{|c|c|c|c|c|c|c|c|c|c|c|c|}
\hline
$a_x$ & $m_{1x}$ & $\Gamma_{1x}$
& $b_{x}$ & $m_{2x}$ & $\Gamma_{2x}$ & $\psi_{2x}^\circ$
& $c_{x}$ & $m_{3x}$ & $\Gamma_{3x}$ & $\psi_{3x}^\circ$
& $\chi^2$ \\
\hline
1.0535 & 782.62 & 8.3768 & 0.30323 & 1019.4 & 4.5055 & \mbox{30.931} &
 3.1673 & 1199.7 & 100.75 & \mbox{-72.319} &
 125.969 \\
\hline
1.0508 & 782.62 & 8.3768 & 0.23505 & 1019.4 & 4.5055 & \mbox{-144.655} &
 3.0147 & 1199.7 & 100.75 & \mbox{-53.143} &
 125.969 \\
\hline
0.99805 & 782.62 & 8.3768 & 0.30038 & 1019.4 & 4.5055 & 153.925 &
3.0128 & 1199.7 & 100.75 & 28.634 & 125.969 \\
\hline
0.99545 & 782.62 & 8.3768 & 0.23284 & 1019.4 & 4.5055 & \mbox{-21.661} &
2.8676 & 1199.7 & 100.75 & 47.811 & 125.969 \\
\hline
\end{tabular}
\end{center}
\end{table}
There are four minimum points with the same values of mass, width
and likelihood function value. It is quite a surprize that the second
resonance phase value are different for all points.
It means that we should see the four minima at the plot of likelihood function,
but we have only two of them.
In principle it can be.
For every new minimization run we take as
a starting point the final point of the previous mi\-ni\-mi\-za\-tion.
Thus the minimization could converge to ``bad'' local minimum.
Let us try to get another plot of likelihood function,
starting minimization at every point  $\psi_2$ closer to the known
``good'' minima (Fig.\,\ref{Fit9_results}).

Now there are all four minimum points on the plot.
However the curve is not smooth, so probably not at every point of  $\psi_{2x}$
the global minimum was achieved, although after covergence
MINUIT executed command {\em IMPROVE},
which tries to seach better minimum.

Let us look which set of minima we can obtain if the resonance width
depends on energy:
\begin{equation}\label{eq:TenthExperiment}
\begin{array}{l}
\sigma(E)=\frac{m_1^4\sqrt{E^2-4\mu^2}^3}{E^4\sqrt{m_1^2-4\mu^2}^3}\times
\left|\frac{2m_1a}{E^2-m_1^2+i{\Gamma_1}{m_1}
\cdot\left(\frac{E^2-4\mu^2}{m_1^2-4\mu^2}
\right)^{\frac{3}{4}}}+
\right.\\ \rule{10mm}{0mm}\left.+
\frac{2m_2be^{i\psi_2}}{E^2-m_2^2+i{\Gamma_2}{m_2}
\cdot\left(\frac{E^2-4\mu^2}{m_2^2-4\mu^2}
\right)^{\frac{3}{4}}}+
\frac{2m_3ce^{i\psi_3}}{E^2-m_3^2+i{\Gamma_3}{m_3}
\cdot\left(\frac{E^2-4\mu^2}{m_3^2-4\mu^2}
\right)^{\frac{3}{4}}}\right|^2
\end{array}
\end{equation}

Again we get the result that for energy dependent resonance width
degeneration disappears (Table~\ref{List_ofScanned_min2}).
\begin{table}[tbp]
\caption{\label{List_ofScanned_min2}Resonance parameters
at the local minima of likelihood function.}
\begin{center}\footnotesize$\rule{0mm}{1mm}\!\!\!\!\!\!\!$%
\begin{tabular}
{|c|c|c|c|c|c|c|c|c|c|c|c|}
\hline
$a_x$ & $m_{1x}$ & $\Gamma_{1x}$
& $b_{x}$ & $m_{2x}$ & $\Gamma_{2x}$ & $\psi_{2x}^\circ$
& $c_{x}$ & $m_{3x}$ & $\Gamma_{3x}$ & $\psi_{3x}^\circ$
& $\chi^2$ \\
\hline
1.0497 & 782.62 & 8.3787 & 0.30316 & 1019.4 & 4.5061 & \mbox{39.985} &
 3.1635 & 1199.7 & 100.79 & \mbox{-64.366} &
 127.005 \\
\hline
1.0470 & 782.62 & 8.3790 & 0.23583 & 1019.4 & 4.5058 & \mbox{-148.442} &
 3.0063 & 1199.7 & 100.71 & \mbox{-45.063} &
 126.529 \\
\hline
0.99824 & 782.62 & 8.3792 & 0.30038 & 1019.4 & 4.5057 & 153.914 &
3.0122 & 1199.7 & 100.71 & 28.687 & 126.522 \\
\hline
0.99559 & 782.62 & 8.3795 & 0.23365 & 1019.4 & 4.5054 & \mbox{-34.675} &
2.8630 & 1199.8 & 100.65 & 47.871 & 126.218 \\
\hline
\end{tabular}
\end{center}
\end{table}

\begin{figure}[tbp]
\epsfxsize=0.48\textwidth
\epsfbox{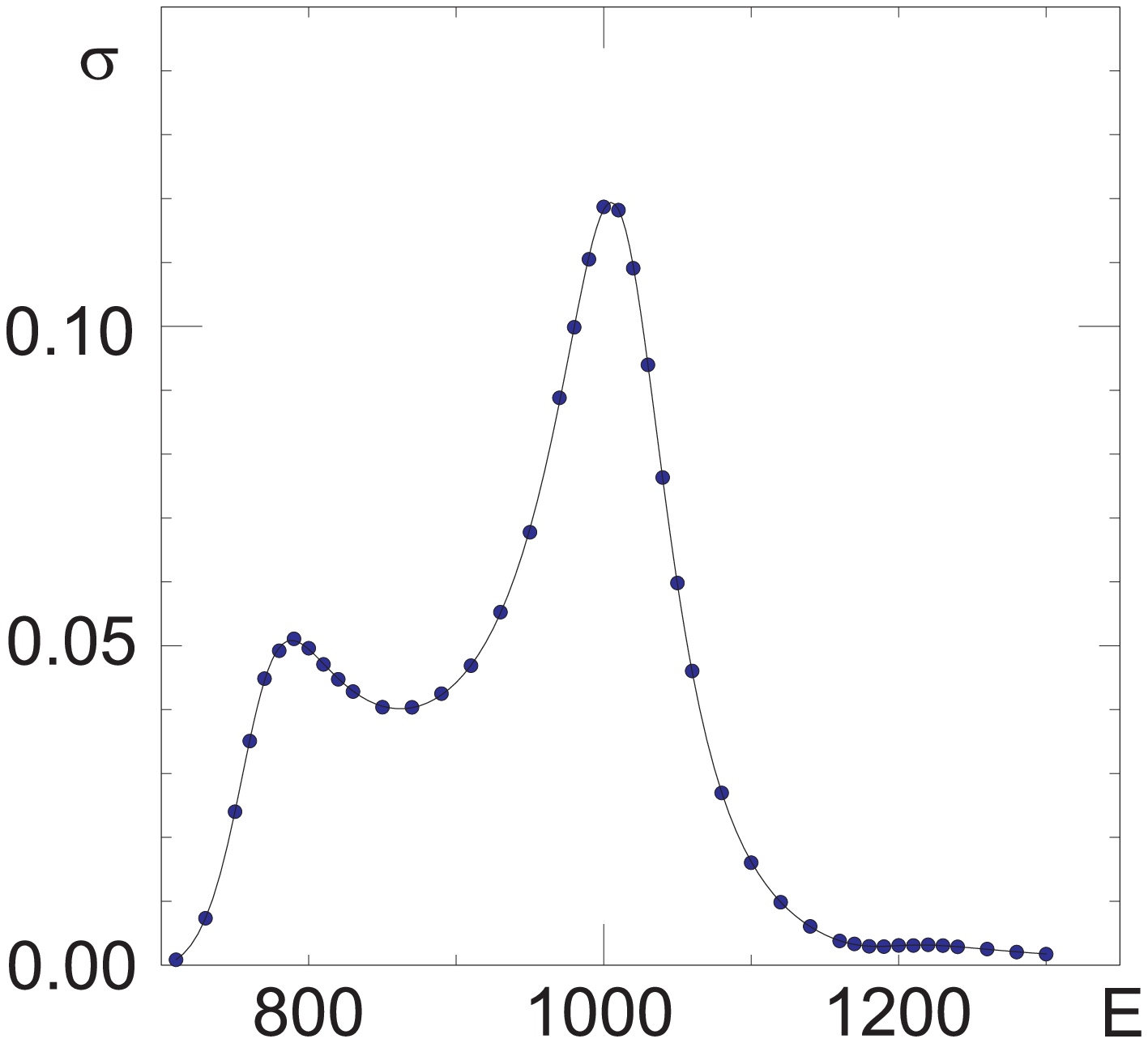}
\hfill
\epsfxsize=0.48\textwidth
\epsfbox{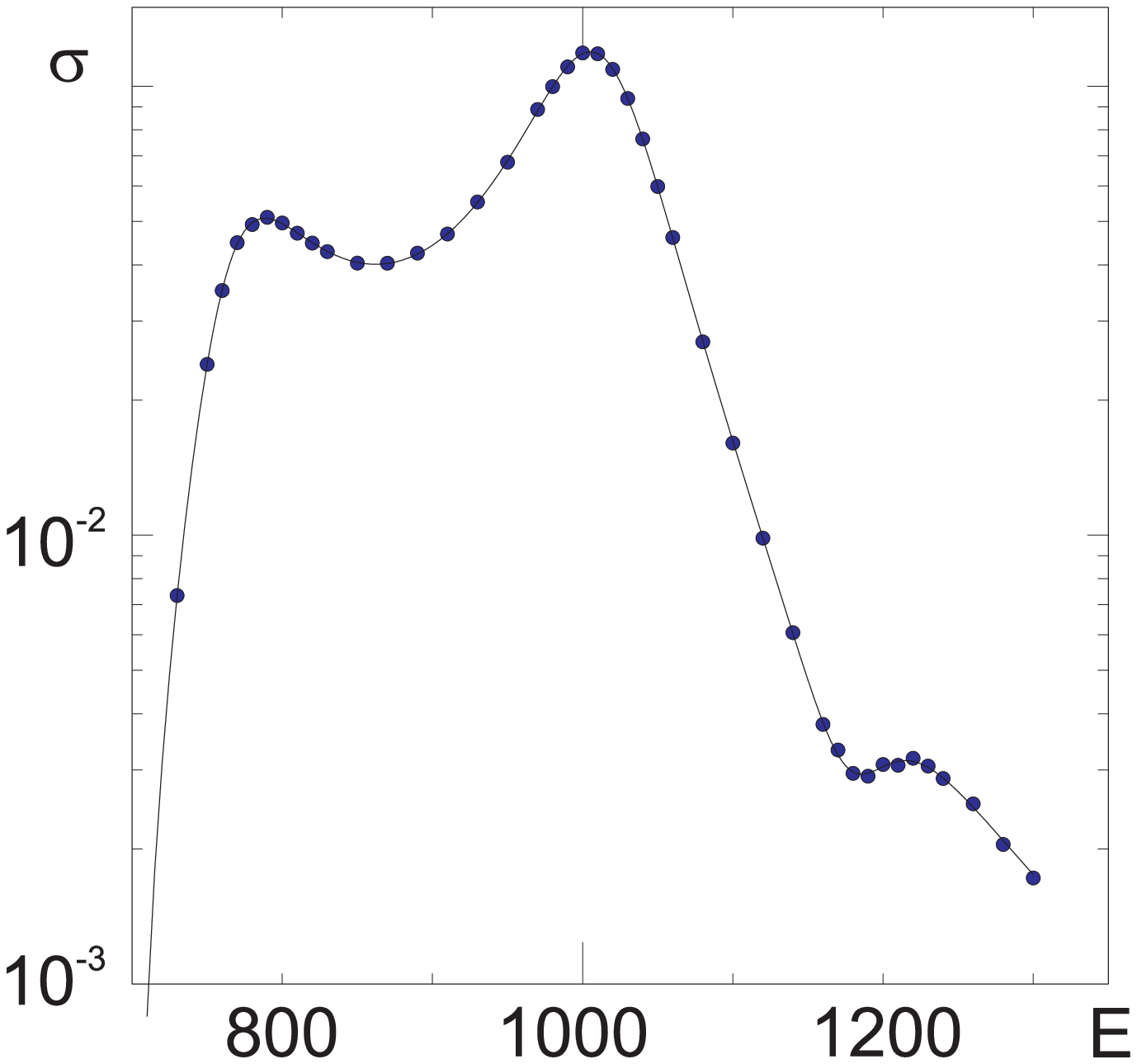}
\caption{\label{FitExp11}Result of fit the ``experimental'' points.
Process cross section is described by the formula~(\ref{eq:TenthExperiment}).
$\chi^2/n_D=37.7/(43-11)$.}
\end{figure}

Again for narrow resonances this difference is negligible from
statistical point of view.
Let us change the following parameters of two resonances:
\begin{equation}
\Gamma_1=100,\;\; \Gamma_2=100,\;\; a=10,\;\; b=10.
\end{equation}
``Experimental'' data and fit result are shown in Fig.\,\ref{FitExp11}.

\begin{figure}[tbp]
\epsfxsize=0.47\textwidth
\centerline{\epsfbox{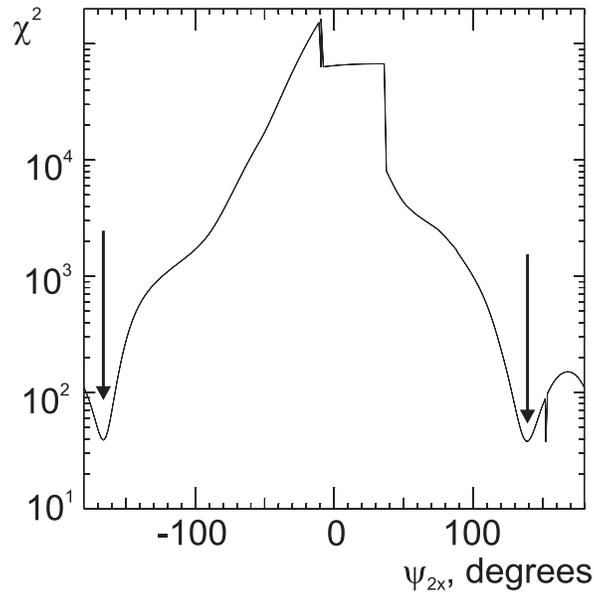}}
\caption{\label{Fit11_results}Likelihood function plot
on the
second resonance phase $\psi_{2x}$
}
\end{figure}

On the likelihood function plot vs  $\psi_{2x}$
(Fig.\,\ref{Fit11_results})
one can see the minima, listed in Table\,\ref{Fit11_2min}.
\begin{table}[tbp]
\caption{\label{Fit11_2min}Resonance parameters at the minimum points
of likelihood function.}
\begin{center}\footnotesize$\rule{0mm}{1mm}\!\!\!\!\!\!\!$%
\begin{tabular}
{|c|c|c|c|c|c|c|c|c|c|c|c|}
\hline
$a_x$ & $m_{1x}$ & $\Gamma_{1x}$
& $b_{x}$ & $m_{2x}$ & $\Gamma_{2x}$ & $\psi_{2x}^\circ$
& $c_{x}$ & $m_{3x}$ & $\Gamma_{3x}$ & $\psi_{3x}^\circ$
& $\chi^2$ \\
\hline
11.685 & 783.30 & 99.681 & 10.503 & 1019.2 & 100.61 & \mbox{-166.020} &
0.75199 & 1199.5 & 104.28 & 157.901 & 38.983 \\
\hline
9.9572 & 782.86 & 100.31 & 10.013 & 1019.0 & 100.10 & \mbox{152.298} &
 3.138 & 1195.7 & 102.44 & \mbox{25.172} &
 37.682 \\
\hline
\end{tabular}
\end{center}
\end{table}

Table\,\ref{PsiScan3} lists the local minimum points found by scanning
angles $\psi_{2x}$  and $\psi_{3x}$.
\begin{table}[tbp]
\caption{\label{PsiScan3}Resonance parameters at the minimum points
of likelihood function.}
\begin{center}\footnotesize$\rule{0mm}{1mm}\!\!\!\!\!\!\!$%
\begin{tabular}
{|c|c|c|c|c|c|c|c|c|c|c|c|}
\hline
$a_x$ & $m_{1x}$ & $\Gamma_{1x}$
& $b_{x}$ & $m_{2x}$ & $\Gamma_{2x}$ & $\psi_{2x}^\circ$
& $c_{x}$ & $m_{3x}$ & $\Gamma_{3x}$ & $\psi_{3x}^\circ$
& $\chi^2$ \\
\hline
9.9572 & 782.86 & 100.31 & 10.013 & 1019.0 & 100.10 & \mbox{152.298} &
 3.138 & 1195.7 & 102.45 & \mbox{25.172} &
 37.682 \\
\hline
9.769 & 782.84 & 100.34 & 8.8903 & 1019.0 & 100.08 & \mbox{138.747} &
0.64948 & 1195.5 & 102.39 & \mbox{68.199} & 37.964 \\
\hline
9.7644 & 782.48 & 99.576 & 8.9027 & 1019.2 & 99.970 & \mbox{141.698} &
0.62815 & 1198.7 & 100.53 & \mbox{77.349} & 41.951 \\
\hline
11.881 & 783.31 & 99.608 & 11.810 & 1019.2 & 100.63 & \mbox{-153.260} &
3.5041 & 1199.8 & 104.31 & \mbox{108.19} & 39.608 \\
\hline
11.685 & 783.30 & 99.681 & 10.503 & 1019.2 & 100.61 & \mbox{-166.026} &
0.75197 & 1199.5 & 104.28 & \mbox{157.901} & 38.983 \\
\hline
\end{tabular}
\end{center}
\end{table}

During this scan five local minima were found. The difference between
lowest minimum and ``highest'' one is significant --- $\Delta\chi^2=4.27$.
However the difference between global minimum and closest one is not so big~---
$\Delta\chi^2=0.28$.
Statistically these two minima are
almost equivalent. Nevertheless the global minimum has
the resonance parameters closer to the ``true'' ones.

\section{Conclusion}

As a result of analytical solution of the problem of parameter
definition of the two interfering resonances by experimental data
there was demonstrated that for cross section parameterization
with constant widths there are always two different solutions
(for different sets of resonance parameters one gets the same
cross section function of energy).
If the dependence of resonance width on energy is taken
into account, then the degeneration disappears, but
quantitavely two solutions usually differ very little,
and this difference is determined by many factors.

For illustration of analytical conclusions a series
of numerical experiments was carried out.
Above conclusions for the case of two
resonances are confirmed, although the statistical
difference of two solutions is not large even if energy
dependence of resonance width is taken
into account.
In every particular case this problem should be investigated
separately.

In the case of three resonances for constant widths
there occurred already four equivalent solutions with
the same likelihood function minimum.
Analytical solution of this problem appeared too hard
due to technical difficulties.
For one numeric example the system of equations was
solved, and four different solutions were derived
(see appendix~\ref{sec:manyRes}).
One can guess that the number of different solutions
equals  $2^{n-1}$, where $n$ is the number of resonances.
In case of any number of resonances the degeneration
disappears when the dependence of resonance
width on energy is taken into account, however
for narrow resonances the statistical difference
between different solutions is usually not significant.

The author is grateful to A.A.\,Korol for his remarks and recommendations.

\bigskip
\clearpage

\appendix
\section{\label{sec:manyRes}Number of solutions for $n$ resonances}
Let us consider a case of $n$ interfering resonances:
\begin{equation}
\sigma(s)=\left|\sum\limits_{k=1}^n\frac{A_k}{s-m_k^2+i\Gamma_km_k}\right|^2,
\end{equation}
where $m_k$, $\Gamma_k$ are mass and width of $k$-th resonance, $m_k<m_{k+1}$,
and $A_k$ are some complex numbers.

This function is entirely defined by the location and residues of its
irregular points, so some other function of the form
\begin{equation}
\sigma_x(s)=\left|\sum\limits_{k=1}^n\frac{A_{kx}}{s-m_k^2+i\Gamma_km_k}\right|^2
\end{equation}
can be equal to the first function over all region of $s$
only if the system of equations is satisfied
\begin{equation}\label{eq:GeneralEquation}
\sum\limits_{k=1}^n\frac{A_{jx}^*\cdot A_{kx}}{m_j^2+i\Gamma_jm_j-m_k^2+i\Gamma_km_k}=
\sum\limits_{k=1}^n\frac{A_{j}^*\cdot A_{k}}{m_j^2+i\Gamma_jm_j-m_k^2+i\Gamma_km_k},\;
j=1,\ldots,n
\end{equation}

If we have only two resonances then the system of equations looks like
\begin{equation}
\left\{
\begin{array}{l}
A_{1x}^*A_{1x}+\frac{2i\Gamma_1m_1}{G}A_{1x}^*A_{2x}
=
A_{1}^*A_{1}+\frac{2i\Gamma_1m_1}{G}A_{1}^*A_{2}
,\\
A_{2x}^*A_{2x}-\frac{2i\Gamma_2m_2}{G^*}A_{2x}^*A_{1x}
=
A_{2}^*A_{2}-\frac{2i\Gamma_2m_2}{G^*}A_{2}^*A_{1}
,
\end{array}
\right.
\end{equation}
where $G=m_1^2+i\Gamma_1m_1-m_2^2+i\Gamma_2m_2$.

Let $A_{1x}=A_1z_1$, $R_1^2=z_1^*z_1$, $A_{2x}=A_2z_2z_1$:
\begin{equation}
\left\{
\begin{array}{l}
A_1^*\cdot\left[\left(R_1^2z_2-1\right)A_2\frac{2i\Gamma_1m_1}{G}+\left(R_1^2-1\right)A_1\right]
=0,\\
A_2^*\cdot\left[
\left(R_1^2\left|z_2\right|^2-1\right)A_2-\frac{2im_2\Gamma_2}{G^*}\cdot
\left(R_1^2z_2^*-1\right)A_1
\right]=0.
\end{array}
\right.
\end{equation}
Now we can derive $z_2$ value from the first equation and substitute to the
second one:
\begin{equation}
\left\{
\begin{array}{l}
z_2=\left(R_1^2-1\right)\frac{iGA_1}{2m_1\Gamma_1R_1^2A_2}+\frac{1}{R_1^2},\\
\left(R_1^2-1\right)\cdot\left(R_1^2-
\frac{1+\frac{2im_1\Gamma_1A_2}{A_1G}-\frac{2im_1\Gamma_1A_2^*}{A_1^*G^*}
+\frac{4m_1^2\Gamma_1^2\left|A_2\right|^2}{\left|A_1\right|^2GG*}}
{1-\frac{4m_1m_2\Gamma_1\Gamma_2}{GG^*}
}
\right)=0.
\end{array}\right.
\end{equation}
One can see that there are two solutions: the first one is
trivial $R_1^2=1$, $z_2=1$, $A_{1x}=A_1$, $A_{2x}=A_2$,
and another solution is
\begin{equation}
R_1^2=\frac{1+\frac{2im_1\Gamma_1A_2}{A_1G}-\frac{2im_1\Gamma_1A_2^*}{A_1^*G^*}
+\frac{4m_1^2\Gamma_1^2\left|A_2\right|^2}{\left|A_1\right|^2GG*}}
{1-\frac{4m_1m_2\Gamma_1\Gamma_2}{GG^*}
}
\end{equation}

If we take the parameters of resonances from the first line of Table~\ref{Fit2_2min}:
\begin{equation}
\begin{array}{lll}
m_1=782.60, & \Gamma_1=8.4116, & A_1=2m_1a=1568.3304,\\
m_2=1019.4, & \Gamma_2=4.5093, & A_2=2m_2be^{i\psi_2}=-555.7337+255.8088i,
\end{array}
\end{equation}
then the second solution should correspond to the second line
in this Table:
$R_1^2=1.010302$, $A_{1x}=A_1R_1=1576.3879$, $z_2=0.6557+0.77886i$, $A_{2x}=A_2R_1z_2=
-563.6463-265.09954i$,
$a_x=\left|A_{1x}\right|/(2m_1)=1.00715$,
$b_x=\left|A_{2x}\right|/(2m_2)=0.3055$, $\psi_{2x}=\arg(A_{2x})=-154.81^\circ$.
Analytical solution matches numerical one within the accuracy
defined by rounding errors.

Now let us carry out similar procedure in case of three resonances
with parameters (the first row in Table~\ref{List_ofScanned_min1}):
\begin{equation}
\begin{array}{lll}
m_1=782.62, & \Gamma_1=8.3768, & A_1=1.0535\cdot 2m_1=1648.98 ,\\
m_2=1019.4, & \Gamma_2=4.5055, & A_2=0.30323\cdot 2m_2e^{i30.931^\circ}=530.3056+317.771i ,\\
m_3=1199.7, & \Gamma_3=100.75, & A_3=3.1673\cdot 2m_3e^{-i72.319^\circ}=2308.1347-7240.6307i
\end{array}
\end{equation}
The system of equations:
\begin{equation}
\left\{
\begin{array}{l}
\left(21.4572-33.7750i\right)R_1^2z_2-\left(224.715-111.758i\right)R_1^2z_3=
3271.14-145.533i-
\\ \rule{0.7\textwidth}{0mm}-
3474.402R_1^2,\\
374.929R_1^2\left|z_2\right|^2-\left(100.699+7.32447i\right)R_1^2z_2^*z_3
+\left(11.5409+18.1660i\right)R_1^2z_2^*=
\\ \rule{0.6\textwidth}{0mm}=
285.770+10.8416i,\\ 48140.6R_1^2\left|z_3\right|^2
-\left(2251.80-163.787i\right)R_1^2z_3^*z_2
-\left(2702.71-1344.14i\right)R_1^2z_3^*=
\\ \rule{0.6\textwidth}{0mm}=
43186.0+1507.93i
\end{array}\right.
\end{equation}
where $R_1^2=\left|z_1\right|^2$.
We can derive variable $z_2$ from the first equation:
\begin{equation}\rule{0mm}{1mm}\!\!\!\!\!\!\!\!\!%
\left\{
\begin{array}{l}
z_2=\left(0.653988+6.23783i\right)z_3-46.5605-73.2892i+\frac{46.9065+67.0514i}{R_1^2},\\
\left(14637.5+623.355i\right)R_1^2\left|z_3\right|^2
-\left(1827002-90862.5i\right)R_1^2z_3^*
-\\ \rule{0.2\textwidth}{0mm}
-\left(177596+97961.8i\right)R_1^2z_3
+\left(168317-93261.3i\right)z_3^*
+\\ \rule{0.2\textwidth}{0mm}
+\left(163103+99669.8i\right)z_3
=\\ \rule{0.3\textwidth}{0mm}
=5321121-67.4339i-2824793R_1^2
-\frac{2510567}{R_1^2},\\
\left(45646.2-13939.2i\right)R_1^2\left|z_3\right|^2
+\left(114146+158750i\right)R_1^2z_3^*
-\left(116606+143303i\right)z_3^*
=\\ \rule{0.3\textwidth}{0mm}
=43186.0+1507.93i
\end{array}\right.
\end{equation}
We got the system of two equations for $z_3$, $z_3^*$, but both equations
are quadratic. Let us introduce $R_3^2=\left|z_3\right|^2$.
From the last equation $z_3^*$ can be derived as a linear expression of $R_3^2$,
then we substitute $z_3$ and $z_3^*$ to another equation and
derive the only root of $R_3^2$:
\begin{equation}
\left\{
\begin{array}{l}
R_3^2=\frac{3705.61R_1^8-13991.2R_1^6+19810.0R_1^4
-12464.7R_1^2+2940.37}{R_1^4\cdot\left(R_1^4-0.892805\right)},\\
z_3=\frac{-\left(290.536+856.587i\right)R_1^6+\left(871.330+2410.08i\right)R_1^4
-\left(867.126+2257.00i\right)R_1^2+286.440+703.515i}{R_1^2\cdot\left(R_1^4-0.892805\right)}.
\end{array}\right.
\end{equation}

Now we can use expression $R_3^2=z_3z_3^*$ as a final
equation for $R_1^2$:
\begin{equation}
\begin{array}{l}
\left(R_1^4-0.94488^2\right)\left(R_1^2-1\right)
\left(R_1^2-0.89281\right)
\left(R_1^2-0.89747\right)
\left(R_1^2-0.99480\right)\times
\\[3mm]\rule{20mm}{0mm}\times
\left[\left(R_1^2-0.94320\right)^2+0.05633^2\right]
\left[\left(R_1^2-0.94447\right)^2+0.02783^2\right]=0
\end{array}
\end{equation}
This is a polynomial of degree 20 in variable $R_1$,
or that of degree 10 in variable $R_1^2$,
so there are 20 formal solutions for $R_1$ or 10 different
solutions for $R_1^2$. But there are only five real roots $R_1^2>0$:
\begin{center}\footnotesize\rule{0mm}{1mm}$\!\!\!\!\!\!$%
\begin{tabular}{|c|c|c|c|c|c|c|c|}
\hline
$R_1$ & $z_2$ & $z_3$ & $a_x=\frac{R_1}{2m_1}$
& $b_x=\frac{\left|A_{2x}\right|}{2m_2}$
& $\begin{array}[t]{l}\psi_{2x}=\\[-1mm]
\mathrm{arctg}\frac{\Im(A_{2x})}{\Re(A_{2x})}\end{array}$
& $c_x=\frac{\left|A_{3x}\right|}{2m_3}$
& $\begin{array}[t]{l}\psi_{3x}=\\[-1mm]
\mathrm{arctg}\frac{\Im(A_{3x})}{\Re(A_{3x})}\end{array}$ \\
\hline
1 & 1& 1& 1.0535 & 0.30323 & $30.931^\circ$ & 3.1673 & $-72.319^\circ$ \\
\hline
0.97205 & $(2.6+5.2i)\cdot 10^{11}$ & $(8.8-3.2i)\cdot 10^{10}$
& 1.0241 & $1.7\cdot 10^{11}$ & $94.5^\circ$ & $2.9\cdot 10^{11}$ & $-92.8^\circ$ \\
\hline
0.9449 & $0.4937-0.6455i$ & $-0.4810+0.8288i$
& 0.99543 & 0.23284 & $-21.663^\circ$ & 2.8677 & $47.810^\circ$ \\
\hline
0.9473 & $-0.5694+0.8771i$ & $-0.1907+0.9858i$
& 0.99803 & 0.30038 & $153.92^\circ$ & 3.0128 & $28.632^\circ$ \\
\hline
0.9974 & $-0.7749-0.0598i$ & $0.9014+0.3135i$
& 1.05076 & 0.23505 & $-144.65^\circ$ & 3.0147 & $-53.14^\circ$ \\
\hline
\end{tabular}
\end{center}
Four of these solutions match with the parameters of resonances
in Table~\ref{List_ofScanned_min1}, and one is very strange
(second row). If we substitute the found solutions to the initial
system of equations, then  the four ``legal'' solutions satisfy
the equations within rounding errors, and ``illegal'' second
solution does not satisfy neither second equation nor the third one.
Obviously this false solution corresponds to the case $z_2=z_3=\infty$,
which should be denied.


Let us consider the case of three resonances where one of them has
infinite width:
\begin{equation}\label{eq:interference_with_constant}
\sigma_x(s)=\left|A_{0x}+\sum\limits_{k=1}^2\frac{A_{kx}}{s-m_k^2+i\Gamma_km_k}\right|^2
\end{equation}
The system of equations reads:
\begin{equation}
\begin{array}{rl}
\left|A_{0x}\right|^2=&\left|A_{0}\right|^2,\\
A_{1x}^*\cdot\left[A_{0x}+\sum\limits_{k=1}^2\frac{A_{kx}}{m_1^2-m_k^2+i\Gamma_km_k
+i\Gamma_1m_1}\right]=&
A_{1}^*\cdot\left[A_{0}+\sum\limits_{k=1}^2\frac{A_{k}}{m_1^2-m_k^2+i\Gamma_km_k
+i\Gamma_1m_1}\right],\\
A_{2x}^*\cdot\left[A_{0x}+\sum\limits_{k=1}^2\frac{A_{kx}}{m_2^2-m_k^2+i\Gamma_km_k
+i\Gamma_2m_2}\right]=&
A_{2}^*\cdot\left[A_{0}+\sum\limits_{k=1}^2\frac{A_{k}}{m_2^2-m_k^2+i\Gamma_km_k
+i\Gamma_2m_2}\right].
\end{array}
\end{equation}
Here we can choose $A_{0x}=A_0$ and $A_{0x}^*=A_{0x}$,
 so we get the system of two equations for
two complex variables $A_{1x},A_{2x}$.
Introduce new notations:
\begin{equation}
A_{1x}=A_1z_1,\;\; A_{2x}=A_2z_1z_2,\;\; \rho_1=\left|z_1\right|^2.
\end{equation}
It is very hard job to derive the solution for general case.
So let us try to solve this problem for the following numerical
example:
\begin{equation}
\begin{array}{lll}
A_0=50, \\
m_1=728, & \Gamma_1= 9, & A_1=\cos 15^\circ -i\sin 15^\circ,\\
m_2=1019, & \Gamma_2 =4, & A_2=39\cdot\left(\cos 155^\circ-i\sin 155^\circ\right).
\end{array}
\end{equation}

\begin{figure}[tbp]
\epsfxsize=0.8\textwidth
\centerline{\epsfbox{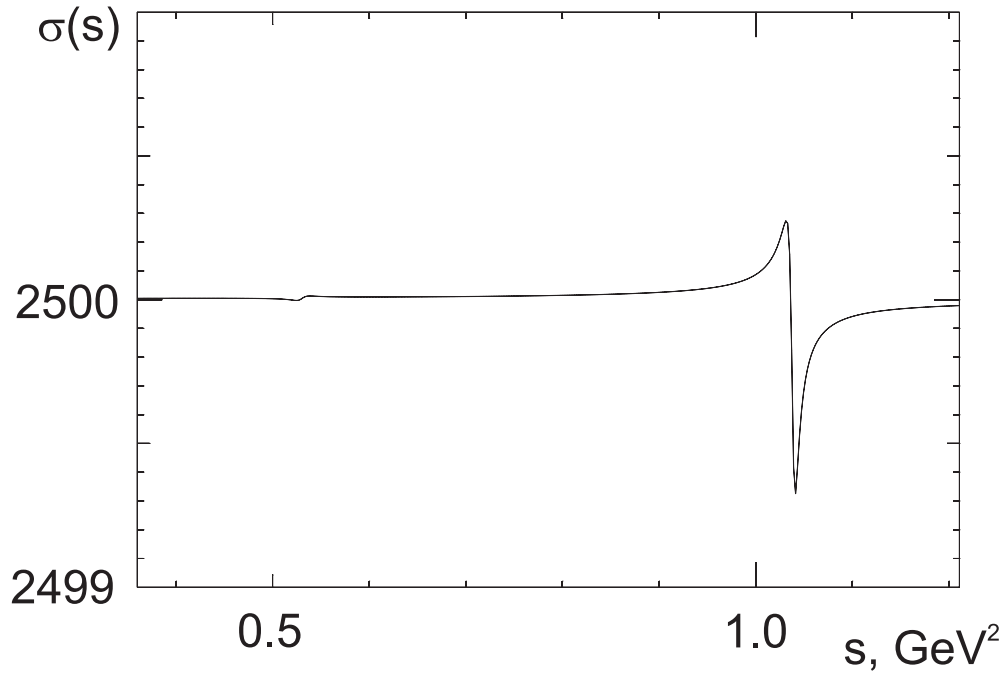}}
\caption{\label{Plot_of_crSec}Plot of the cross section~(\ref{eq:interference_with_constant})
with amplitude parameters from the first row in Table~\ref{tabOfRoots}.}
\end{figure}

\begin{figure}[tbp]
\epsfxsize=0.49\textwidth
\epsfbox{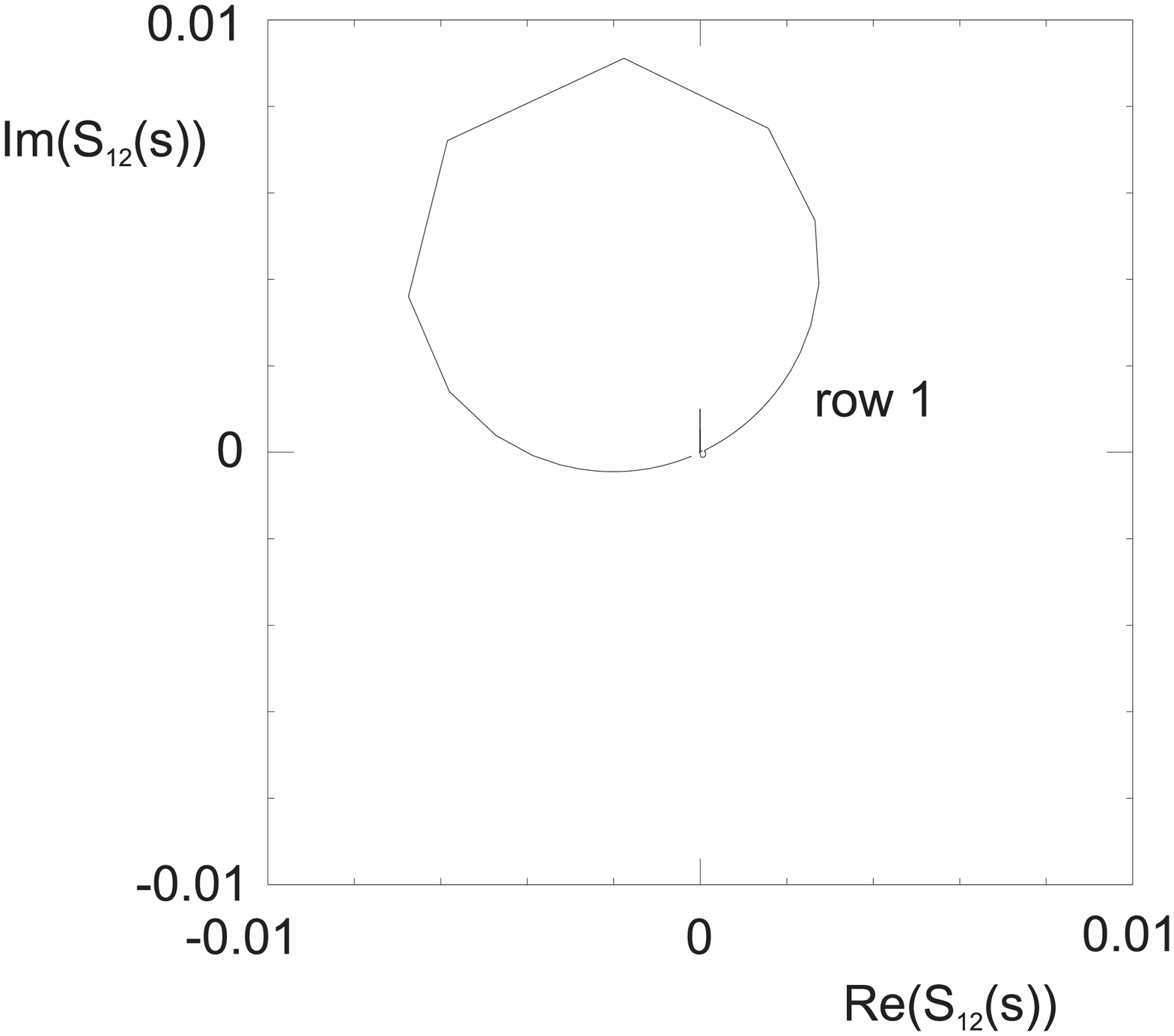}
\hfill
\epsfxsize=0.49\textwidth
\epsfbox{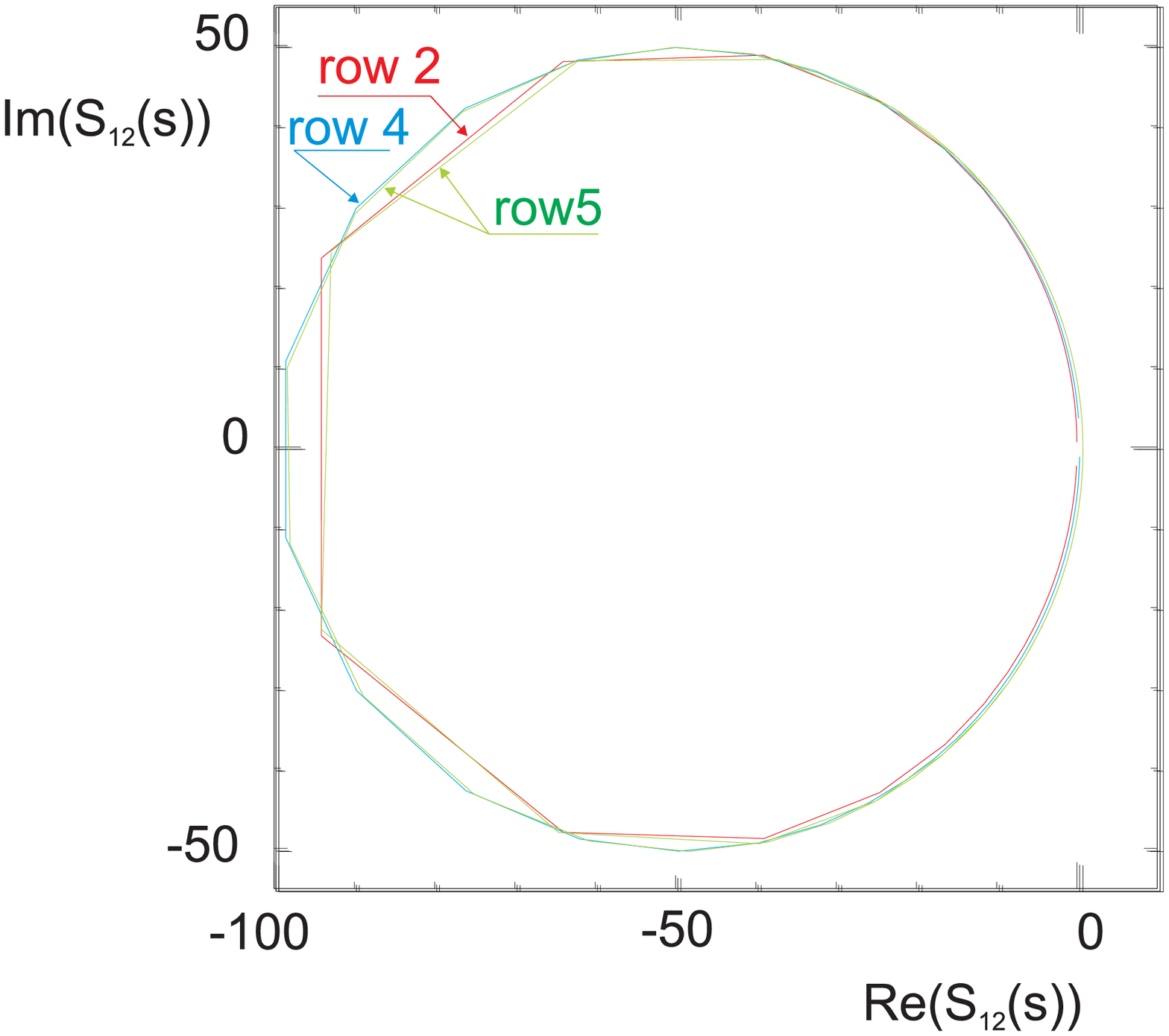}
\caption{\label{TrajectoriesS12} Trajectories $S_{12}(s)$ for different solutions
(marked by the number of row in Table~\ref{tabOfRoots}) for the parameter
$s\in (0.36,1.21)$ GeV$^2$. Left picture for the solution in the first row,
right picture is for all the rest solutions. Step for the trajectory plot equals
$\Delta \sqrt{s}=1$ MeV.}
\end{figure}

$z_1,z_2$ can be re-written as functions of $\rho_1$, and for $\rho_1$ we get the
algebraic equation of 12-th degree:
\begin{equation}\label{eq:FinalEquation}
\begin{array}{l}
\rule{3mm}{0mm}\left(\rho_1-1\right)\cdot\left(
-3.743751\cdot 10^{67}+
\left(\rho_1-1\right)\cdot\left(
9.058522\cdot 10^{70}+ \right.\right.
\\ +\left(\rho_1-1\right)\cdot\left(
1.511509\cdot 10^{65}+
\left(\rho_1-1\right)\cdot\left(
-1.333590\cdot 10^{59}+
\right.\right.
\\ +\left(\rho_1-1\right)\cdot\left(
-5.218886\cdot 10^{53}+
\left(\rho_1-1\right)\cdot\left(
-8.183609\cdot 10^{47}+
\right.\right.
\\ +\left(\rho_1-1\right)\cdot\left(
-2.813253\cdot 10^{40}+
\left(\rho_1-1\right)\cdot\left(
1.405807\cdot 10^{36}+
\right.\right.
\\ +\left(\rho_1-1\right)\cdot\left(
9.868724\cdot 10^{29}+
\left(\rho_1-1\right)\cdot\left(
1.843597\cdot 10^{23}+
\right.\right.
\\
\left.\left.\left.\left.\left.\left.\left.\left.\left.\left.
\hspace{-4mm}+\left(\rho_1-1\right)\cdot\left(
-8.587479\cdot 10^{11}+
\left(\rho_1-1\right)
\right)
\right)\right)\right)\right)\right)\right)\right)\right)\right)
\right)=0.
\end{array}
\end{equation}
All roots (both real and complex) are located within circle $\left|\rho_1\right|=10^{13}$.
 One root is trivial: $\rho_1=1,
z_1=z_2=1$. Using Sturm method~\cite{KORN} for the remaining polynomial
of 11-th order, there was found, that some roots are doubled,
and the polynomial
\begin{equation}
(\rho_1-1)^3+2.49\cdot 10^6(\rho_1-1)^2-4.29\cdot 10^{11}(\rho_1-1)
-1.07\cdot 10^{18}
\end{equation}
was a common divisor for the original polynomial and its derivative.
Repeating the Sturm procedure for the remaining polynomial of
8-th order,
it is possible to check that there are 6 real roots of this equation.
Preliminary localization defined exactly one root between
every two of the following points:
\begin{equation}
-4.9\cdot 10^{6}, -2.5\cdot 10^6, 0, 4.1\cdot 10^5, 8.2\cdot 10^5,
4.295\cdot 10^{11}, 4.297\cdot 10^{11}.
\end{equation}
Negative roots should be rejected, because $\rho_1$ can be only positive.
Table~\ref{tabOfRoots} presents the values of positive roots and some
additional information.
\begin{table}[tbp]
\caption{\label{tabOfRoots}Real and positive roots of the equation~(\ref{eq:FinalEquation}).}
\begin{tabular}{|c||c|c|c||c|c|c|}
\hline
$\rho_1$ & $A_{1x}$ & $\left|A_{1x}\right|$ & $\psi_1$, deg. &
 $A_{2x}$ & $\left|A_{2x}\right|$ & $\psi_2$, deg. \\
\hline
1.000000 &$0.9659+0.2588i$ &1.0000 &-15.00 &$-35.346-16.482i$ &39.000 &-155.00 \\
\hline
1.000413 &$0.9702-0.2434i$ &1.0002 &-14.08 &$-35.350-4.08\cdot 10^5i$ &$4.076\cdot 10^5$ &-90.005 \\
\hline
655336.0 &$\infty$ & & &$\infty$ & & \\
\hline
$4.292879\cdot 10^{11}$ &$1.3863-6.552\cdot 10^5i$ &$6.552\cdot 10^5$ &-90.000 &
  $-35.766-15.569i$ &39.008 & -156.48 \\
\hline
$4.294653\cdot 10^{11}$ &$10507-6.552\cdot 10^5i$&$6.553\cdot 10^5$ &-89.081 &
 $-10541-4.075\cdot 10^5i$ &$4.077\cdot 10^5$ &-91.482 \\
\hline
\end{tabular}
\end{table}
The third row in this table has inappropriate solution,
because $z_1$ for it goes to infinity.
$z_1$ is the ratio of two polynomials and for $\rho_1=655336$ polynomial in
the denominator equals zero.
The last two solution are quite unexpected because of high value of amplitudes.
Original cross section is shown in Fig.\,\ref{Plot_of_crSec}.
Within the same interval of $s$ the ratio of cross section of alternative
solution and the original cross section was evaluated and
occurred to be equal to 1 with high accuracy.
In order to avoid some digital surprises
 all these calculations were carried out with high accuracy
of 150 decimal digits, using REDUCE system~\cite{REDUCE}.
In order to illustrate the strange two last solutions, in Fig\,\ref{TrajectoriesS12}
the trajectories of the complex function
\begin{equation}
S_{12}(s)=\frac{A_{1x}}{s-m_1^2+i\Gamma_1m_1}+\frac{A_{2x}}{s-m_2^2+i\Gamma_2m_2}
\end{equation}
are presented on the complex plane.


So we got two solutions for the case of two resonances and four solutions
for the case of three resonances.
It is not enough to choose the rule for the number $N_s$ of solutions
for $n$ resonances: it can be $N_s=2^{n-1}$, or $N_s=2(n-1)$,
or something else.

In order to check whether it is possible to solve the
system of equations in every case (at least numerically),
let us solve a similar problem with four resonances,
but choose the most simple input data making easier all
calculations:
\begin{equation}
\begin{array}{lll}
m_1=1, & \Gamma_1=1, & A_1= 1 ,\\
m_2=2, & \Gamma_2=1, & A_2= 1 + i, \\
m_3=3, & \Gamma_3=1, & A_3= 1 - i, \\
m_4=4, & \Gamma_4=1, & A_4= -1 + i. \\
\end{array}
\end{equation}
 The system of equations looks like
\begin{equation}\label{eq:MainSystem}\left\{
\begin{array}{l}
A_{x1}^*\cdot\left(A_{x1}+\frac{1-i}{3}A_{x2}+\frac{1-2i}{10}A_{x3}
+\frac{1-3i}{25}A_{x4}\right)=\frac{247-21i}{150},\\
A_{x2}^*\cdot\left(\frac{2+2i}{3}A_{x1}+A_{x2}+
\frac{2-2i}{5}A_{x3}+\frac{2-4i}{15}A_{x4}\right)=\frac{46-8i}{15},\\
A_{x3}^*\cdot\left(\frac{3+6i}{10}A_{x1}+\frac{3+3i}{5}A_{x2}
+A_{x3}+\frac{3-3i}{7}A_{x4}
\right)=\frac{207i-25}{70},\\
A_{4x}^*\cdot\left(\frac{4+12i}{25}A_{x1}+\frac{4+8i}{15}A_{x2}
+\frac{4+4i}{7}A_{x3}+A_{x4}
\right)=\frac{1178-1216i}{525}.
\end{array}\right.
\end{equation}

If we describe the j-th equation in the form
\begin{equation}
A_{xj}^*\cdot\sum\limits_{k=1}^4G_{jk}A_{xk}=R_j,
\end{equation}
then the solution of every equation can be written as follows
\begin{equation}
A_{xj}=-\frac{1}{2}S_j-\frac{i\Im(R_j)}{S_j^*}+S_jQ_j
=S_j\cdot\left(Q_j-\frac{1}{2}-\frac{i\Im(R_j)}{\left|S_j\right|^2}\right),
\end{equation}
where
\begin{equation}\label{eq:QjAndQsinj}
S_j=\sum\limits_{k\neq j}G_{jk}A_{xk},\;\;
Q_j^2=\frac{1}{4}+\frac{\Re(R_j)}{\left|S_j\right|^2}-\frac{\Im(R_j)^2}{\left|S_j\right|^4}.
\end{equation}

These four solutions together can be considered as a system of linear equations:
\begin{equation}\label{eq:LinearSystem}
\sum\limits_{k=1}^nB_{jk}A_{xk}=0,\;\;
B_{jk}=\left\{
\begin{array}{l}
\frac{1}{C_j},\;\; k=j,\\
G_{jk},\;\; k\neq j,
\end{array}\right. \;\;\;
 j=1,\ldots,n,
\end{equation}
where
\begin{equation}\label{eq:CjDependence}
C_j=\frac{1}{2}+\frac{i\Im(R_j)}{\left|S_j\right|^2}
\pm\sqrt{
\frac{1}{4}+\frac{\Re(R_j)}{\left|S_j\right|^2}-\frac{\Im(R_j)^2}{\left|S_j\right|^4}
}
\end{equation}
This system can have non-zero solution only if the determinant is equal to zero.
If it is, we can use the last three equations to express
all amplitudes $A_{xj}$ through the amplitude $A_{x1}$:
\begin{equation}
\begin{array}{l}
A_{x2}=\frac{3}{5}\cdot\frac{ \left(876+4908i\right)C_3C_4+
\left(6615+2205i\right)C_3+\left(2744+392i\right)C_4-12250(1+i)}
{ 4032C_2C_3C_4-5292C_2C_3-1960C_2C_4-5400C_3C_4+11025}C_2A_{x1},\\[3mm]
A_{x3}=\frac{21}{50}\cdot\frac{ -\left(4616+7888i\right)C_2C_4+
21000iC_2+3600\left(i+2\right)C_4-7875\left(1+2i\right)}
{ 4032C_2C_3C_4-5292C_2C_3-1960C_2C_4-5400C_3C_4+11025}C_3A_{x1},\\[3mm]
A_{x4}=\frac{14}{25}\cdot\frac{ \left(9882-8874i\right)C_2C_3+
\left(3i-1\right)\left(3500C_2-3375C_3\right)-3150\left(1+3i\right)}
{ 4032C_2C_3C_4-5292C_2C_3-1960C_2C_4-5400C_3C_4+11025}C_4A_{x1}.
\end{array}
\end{equation}
If the variables $C_j$ were the predefined constants, then
these expressions would be the set of infinite number of solutions
with arbitrary $A_{x1}$. But here $C_j$ depend on $A_{xk}$
via the relation (\ref{eq:CjDependence}). And even more, instead of
constraint on the determinant of the system~(\ref{eq:LinearSystem})
we can use the first equation of the system~(\ref{eq:MainSystem}),
which can be presented in the form:
\begin{equation}
\begin{array}{l}
R_1^2=F(C_2,C_3,C_4)=
\frac{2\left(247-21i\right)}{3}\times\\[3mm] \rule{4mm}{0mm}\times\frac
{4032C_2C_3C_4-5292C_2C_3-1960C_2C_4-5400C_3C_4+11025}
{395736C_2C_3C_4-176400C_2C_3-70560C_2C_4-490000C_2-419040C_3C_4-165375C_3-70560C_4
+1102500},
\end{array}
\end{equation}
where $R_1=\left|A_{x1}\right|$.

The problem looks very much complicated. Let us try to solve it
using numeric minimization procedure.
The free parameters are complex variables $C_2,\,C_3,\,C_4$.
Minimized function
\begin{equation}
\Phi=\left(\Im\left(F(C_2,C_3,C_4)\right)\right)^2+
\sum\limits_{j=2}^4\left|
\frac{1}{2}+\frac{i\Im(R_j)}{\left|S_j\right|^2}
\pm\sqrt{
\frac{1}{4}+\frac{\Re(R_j)}{\left|S_j\right|^2}-\frac{\Im(R_j)^2}{\left|S_j\right|^4}
}-C_j
\right|^2
\end{equation}

The true solution is found if the minimum value of $\Phi$ equals zero.
Eight different combinations of signs of square roots provide possible
eight solutions.
Attempt to use the code MINUIT~\cite{MINUIT} failed because of very complicated
function profile. Use of \cite{BUKMIN} brought more success.
Table \ref{tab:4resonMinima} presents the results of minimization.
\begin{table}[tbp]
\caption{\label{tab:4resonMinima}Results of the search for the
solutions in case of four resonances}
\begin{center}
\begin{tabular}{|c|c|c|c|c|c|c|c|}
\hline
$Q_2$ & $Q_3$ & $Q_4$ & $R_1$ & $A_{x2}$ & $A_{x3}$ & $A_{x4}$ & $\min \Phi$ \\
\hline
-2.000 & -0.1704 &-0.5899 &1.0000 & $1.0000+1.0000i$ & $0.9999-1.0002i$ & $-1.0000+1.0000i$ &
$7.5 \cdot 10^{-9}$ \\
\hline
-2.8900 & $-4\cdot 10^{-8}$ &0.6880 &1.0492 & $0.9378+1.2978i$ & $1.9073-0.3002i$ &
$-2.0751-2.2721i$ &
$0.009$ \\
\hline
$-7\cdot 10^{-6}$ & 0.3203 &-0.3965
&1.0459 & $1.0731+1.3992i$ & $0.5103-2.5446i$ & $-0.4493+1.5012i$ &
$4.2 \cdot 10^{-4}$ \\
\hline
-2.9180 & 0.1103 &0.6893 &1.0536 & $0.9359+1.3216i$ & $1.9127-0.4163i$ & $-2.0985-2.2274i$ &
$8.1\cdot 10^{-8}$ \\
\hline
0.9510 & -0.0139 &-0.5047 &2.2669 & $-1.8270-2.5487i$ & $-1.7284+0.0352i$ & $1.4877+0.2031i$ &
$6.5\cdot 10^{-4}$ \\
\hline
0.8089 & $-5\cdot 10^{-8}$ &0.6846
&2.4065 & $-1.8727-3.2613i$ & $-1.4980-1.0233i$ & $-0.6968+3.0950i$ &
$0.018$ \\
\hline
0.7616 & 0.3213 &-0.4026 &1.0472 & $1.0754+1.4242i$ & $0.4838-2.6127i$ & $-0.3946+1.5174i$ &
$4.6 \cdot 10^{-11}$ \\
\hline
2.5546 & 0.4321 &0.6691 &1.1033 & $0.9392+1.8158i$ & $2.5179-2.6845i$ & $-3.2574-0.9483i$ &
$1.3\cdot 10^{-7}$ \\
\hline
\end{tabular}
\end{center}
\end{table}

Despite minimization problems this algorithm allows to localize the solutions, exactly
$2^{n-1}$ of them, where $n$ is the number of resonances.
To improve the amplitudes values and make sure that localization is good enough,
one can minimize the function
\begin{equation}
\Psi=\sum\limits_{j=1}^4\left|A_{xj}^*\sum\limits_{k=1}^4G_{jk}A_{xk}-R_j\right|^2,
\end{equation}
starting minimization from the found points.
There are 7 free paramerters: $R_1=\Re(A_{x1})$ and real and imaginary parts of
$A_{x2}$, $A_{x3}$, $A_{x4}$ ($\Im(A_{x1})=0$).
The result of this operation is shown in Table~\ref{tab:4resImprov}.
\begin{table}[tbp]
\caption{\label{tab:4resImprov}Improved parameters of solutions.
Minimization of $\Psi$ started from the approximation from the Table\ref{tab:4resonMinima}.}
\begin{minipage}{\textwidth}
\begin{center}
\begin{tabular}{|c|c|c|c|c|}
\hline
$A_{x1}$ & $A_{x2}$ & $A_{x3}$ & $A_{x4}$ & $\min \Psi$ \\
\hline
$1.0000+0i$\footnotemark[1]%
\footnotetext[1]{Improved solution point matched the initial approximation
from the Table~\ref{tab:4resonMinima}}
& $1.0000+1.0000i$ &
       $1.0000-1.0000i$ &
                   $-1.0000+1.0000i$ & $1.2\cdot 10^{-13}$ \\
\hline
$1.0536+0i$\footnote[2]{The minimum point moved avay essentially from the
approximation in Table~\ref{tab:4resonMinima} and matched the solution
in the fourth row}
 & $0.9361+1.3216i$ &
       $1.9126-0.4174i$ &
                   $-2.0986-2.2262i$ & $5.8\cdot 10^{-8}$ \\
\hline
$1.0471+0i$\footnote[3]{The minimum point moved avay essentially from the
approximation in Table~\ref{tab:4resonMinima} and matched the solution
in the seventh row}
 & $1.0762+1.4232i$ &
       $0.4831-2.6116i$ &
                   $-0.3947+1.5171i$ & $6.2\cdot 10^{-6}$ \\
\hline
$1.0536+0i$\footnotemark[1] & $0.9360+1.3217i$ &
       $1.9126-0.4172i$ &
                   $-2.0985-2.2262i$ & $3.9\cdot 10^{-11}$ \\
\hline
$2.2528+0i$\footnotemark[1] & $-1.8210-2.5396i$ &
       $-1.7049+0.5456i$ &
                   $1.4949+0.2602i$ & $1.9\cdot 10^{-6}$ \\
\hline
$2.3744+0i$\footnote[4]{The minimum point moved avay essentially from the
approximation in Table~\ref{tab:4resonMinima}, $Q_2=0.811$, $Q_3=0.253$, $Q_4=0.676$}
 & $-1.5650-3.2183i$ &
       $-2.0232-1.2148i$ &
                   $-0.6648+3.2157i$ & $1.3\cdot 10^{-5}$ \\
\hline
$1.0472+0i$\footnotemark[1] & $1.0754+1.4242i$ &
       $0.4838-2.6127i$ &
                   $-0.3946+1.5174i$ & $9.6\cdot 10^{-15}$ \\
\hline
$1.1033+0i$\footnotemark[1] & $0.9393+1.8152i$ &
       $2.5170-2.6818i$ &
                   $-3.2560-0.9502i$ & $7.5\cdot 10^{-10}$ \\
\hline
\multicolumn{5}{c}{Additionally found solutions}\\
\hline
$2.4858+0i$\footnote[5]{Reconstructed values $Q_2=0.676$, $Q_3=0.461$, $Q_4=0.653$}
 & $-1.3958-4.2950i$ & $-4.4264+0.2797i$ &
       $1.1745+3.4443i$ & $4.8\cdot 10^{-10}$ \\
\hline
$2.3593+0i$\footnote[6]{Reconstructed values $Q_2=0.743$, $Q_3=0.397$, $Q_4=-0.210$}
 & $-1.8330-3.4916i$ & $-2.5821+1.8974i$ &
       $1.5754-0.5896i$ & $4.9\cdot 10^{-9}$ \\
\hline
\end{tabular}
\end{center}
\end{minipage}
\end{table}
One can see that for those cases, where all $Q_j$ had non-zero values,
the improved points practically match the approximate values of $A_{xj}$.
On contrary, for ``bad'' points (rows 2,3 and 6), the improved values
of $A_{xj}$ are rather far from approximation in Table~\ref{tab:4resonMinima}.
Furthermore the found solutions in rows 2 and 3 match exactly other
solutions, so the approximations in Table~\ref{tab:4resonMinima}
were not close to some new solutions.
Starting randomly from different points, one can find additional
two solutions, presented at the bottom of Table~\ref{tab:4resImprov}.

This exercise shows that the suggested algorithm cannot localize reliably
all solutions of this problem. But it supports the rule $2^{n-1}$ for the
number of solutions for $n$ resonances problem.



\vspace{\stretch{1}}

\tableofcontents

\vspace{\stretch{1}}


\begin{thebibliography}{99}
\bibitem{BINP_prep}
{\it A.D.\,Bukin.} On the ambiguity of the parameters of interfering resonances.
Preprint Budker INP 2007-24, Novosibirsk, 2007 (in Russian).\\
URL=http://www.inp.nsk.su/activity/preprints/files/2007\_024.pdf
\bibitem{KORN}
{\it G.Korn and T.Korn.} Mathematical handbook for scientists and
engineers. McGraw-Hill Book Company, 1968
\bibitem{MINUIT}
{\it F.\,James, M.Roos.} 'MINUIT' A System for Function Minimization
and Analysis of the Parameter Errors and Correlations,
Computer Physics Communications 10 (1975) 343. \\
{\it F.\,James.} MINUIT. Function Minimization and Error Analysis.
Reference Manual. CERN Program Library Long Writup D506, March 1994.
\bibitem{BUKMIN}
{\it A.Bukin.} Subroutine for numerical minimization of the function
of many parameters.Preprint BINP 2004-78, Novosibirsk, 2004 (in Russian)
\bibitem{REDUCE}
{\it Anthony C.Hearn.} REDUCE user's manual. Rand, 1987.
\end{thebibliography}
\end{document}